\documentclass{aa} 

\usepackage{graphicx}
\usepackage{epsfig,latexsym,longtable,lscape,supertabular}
\usepackage[authoryear]{natbib}

\bibpunct{(}{)}{;}{a}{}{,} 

\usepackage{txfonts}
\begin{document}
   \title{The diverse X-ray properties of four truly isolated elliptical galaxies: NGC 2954, NGC 6172,
 NGC 7052, and NGC 7785}

\author{E.~Memola\inst{1,*}, G.~Trinchieri\inst{1}, A.~Wolter\inst{1}, P.~Focardi\inst{2}, \and B.~Kelm\inst{2}}

\offprints{E.~Memola,  elisabetta.memola@mib.infn.it
* Currently at INFN-Istituto Nazionale di Fisica Nucleare, Sezione di Milano Bicocca, Piazza della Scienza 3, 20126 Milano, Italy}

\institute{INAF-Istituto Nazionale di Astrofisica, Osservatorio Astronomico di Brera, Via Brera 28, 20121 Milano, Italy 
\and Dipartimento di Astronomia, Universit\`a di Bologna, Via Ranzani 1, 40127 Bologna, Italy 
}

\date{Received 13 August 2008 / Accepted 4 February 2009}

 
  \abstract
   {}
   {We investigate the X-ray properties of four isolated elliptical galaxies,
selected from the Updated Zwicky Catalog according to strict 
isolation criteria.
Isolated galaxies are not influenced by the group/cluster environment,
and their X-ray emission can be studied independently of the often 
overwhelming contribution of the hot intergalactic medium.  They are
therefore suited to studying the X-ray characteristics relative to 
their intrinsic properties. 
}
   {We analyzed our own XMM-{\em Newton} and archival Chandra data in detail
for three objects, and derived, when possible, 
the spatial and spectral characteristics of each source. An upper limit
for the fourth one was obtained from archival ASCA data. 
We compared their characteristics with those of other 
23 isolated objects
for which X-ray and optical data are 
available in the literature. 
We explored possible theoretical explanations to interpret our results. 
}
   {In spite of our attempt to select very homogeneous objects, both in terms of
optical properties and environmental characteristics, we find a
wide range in X-ray luminosities and  L$_{\rm X}$/L$_{\rm B}$ ratios for the
four objects: two of them show a hot gaseous halo, 
whereas no gas is detected in the other two, to a factor $>10$ in
luminosity. In fact, we find a large spread in the 
L$_{\rm X}$/L$_{\rm B}$ for all galaxies considered, 
suggesting that the presence of hot gas is not easily related 
to the optical luminosity or to the  mass, even in isolated
systems. 
Younger objects tend to be less luminous in X-rays than older systems.  
However, it appears that 
older objects could span a wide range in luminosities. 
} 
   {}

\authorrunning{E.~Memola et al.}
\titlerunning{Isolated ellipticals}

   \maketitle
%

\section{Introduction}

The high X-ray luminosity observed in early-type galaxies 
comes mostly from a diffuse hot component  (e.g.~Forman et al.~1985;
Trinchieri \& Fabbiano 1985), 
while stellar sources dominate only at the
low end of the X-ray luminosity distribution
(e.g.~Fabbiano \& Trinchieri 1985; Kim et al.~1992).  
It has long been known that the relation between 
the X-ray and the optical luminosity
of early-type galaxies shows a large scatter
(e.g.~Canizares et al.~1987; Eskridge et al.~1995; 
Brown \& Bregman 1998; Beuing et al.~1999). 
The contribution from a group and its effects on the galaxy X-ray 
properties can explain some of this scatter
(Helsdon et al.~2001). Massive ellipticals at the center of
a group retain their own interstellar medium (ISM) and may also be
embedded in the hot
intra-group gas, which could enhance their observed X-ray luminosity
(Canizares et al.~1983); at the same time, 
a galaxy moving at high speed through the intra-group medium may 
undergo ram-pressure stripping (Gunn \& Gott 1972)
or viscous stripping (Nulsen 1982), thus decreasing
its gaseous content without affecting its stellar luminosity much. 

The majority of early-type galaxies are found in groups
and clusters (Melnick \& Sargent 1977; Dressler 1980; Tully 1987),
and  dominant members
are the most X-ray luminous examples of their type 
(Helsdon et al.~2001; O'Sullivan et al.~2001b).
X-ray selected samples are, thus, most likely  biased towards X-ray bright,
hence group dominant ellipticals. 
The influence  of the environment has therefore
been one of the major sources of confusion, since it is difficult to
disentangle the group and the galaxy properties.
On the other hand, it has already been noticed that the scatter in
the L$_{\rm X}$--L$_{\rm B}$ relation persists 
even when center group galaxies are not considered, and it is relatively 
large at any given mass.
For instance, by excluding galaxies embedded in an intra-group medium 
\citep[e.g.~the EXG in ][]{fukazawa},
part of the scatter 
can be reduced, but only at high luminosities
(i.e.~most massive early-type galaxies). The careful selection of galaxies within an extremely poor environment is a 
step towards quantifying the environmental effects.

Many studies have addressed the issue of defining optical samples of isolated
galaxies (e.g.~Colbert et al.~2001; Smith et al.~2004; Reda et al.~2004; 
Denicol\'o et al.~2005) but unfortunately only a few have a follow-up in the 
X-ray band (e.g.~O'Sullivan \& Ponman 2004; O'Sullivan et al.~2007). 
In this work we have considered a small, local sample selected with
strict isolation criteria and studied its X-ray emission. 
We also intend to collect additional data (e.g.~optical spectra to derive
Lick indexes), to better define the global properties of our isolated ellipticals, 
which will help our understanding of their X-ray characteristics in the 
context of the current formation and evolution scenarios for ellipticals. 
While it is established that the stars in the most
massive ellipticals were formed at
high redshift (Ellis et al.~1997; Treu et al.~2002; Bernardi et al.~2003;
van Dokkum 2005), much  less is known about how massive ellipticals have been
assembled (Andreon 2006; De Propris et al.~2007; De Lucia et al.~2008), 
and current evidence has so far been unable to unambiguously prefer
the monolithic  scenario (e.g.~Eggen et al.~1962;
Larson 1975; Chiosi \& Carraro 2002) versus
hierarchical merging  (e.g.~Toomre \& Toomre 1972; Knochfar \& Silk 2006).
Moreover,  in their study of a sample of isolated ellipticals,
Marcum et al.~(2004) find both good candidates for the  
monolithic assembly scenario and objects that bear clear signs of 
interaction (up to possible fossil groups). 
In the hierarchical merger scenario, mergers should occur at all times, 
and they might be observable in the field in the present epoch. 
Recent works studied the evolution of the X-ray emission along
the merging sequence (Nolan et al.~2004; Brassington et al.~2007), 
which can provide some prediction on the expected X-ray luminosity.
In particular, post-merger ellipticals could be relatively devoid of the
X-ray gas, possibly expelled at the nuclear merger stage, while older
merger remnants could have had the time to rebuild a sizable X-ray halo 
(O'Sullivan et al.~2001a; Sansom et al.~2006).
Observing isolated galaxies would allow us to study these effects
independently of the environmental factor.

We discuss here the X-ray properties for the four early-type galaxies 
from the sample we selected (Section 2) for which X-ray data exist.
Throughout this paper we assume H$_{0}$=75~km~s$^{-1}$~Mpc$^{-1}$.
We normalize optical $B$-band and $K$-band luminosities
to the $B$-band and  $K$-band luminosity of the sun,
L$_{\rm B_\odot} \sim 5 \times 10^{32}$~erg~s$^{-1}$, and  
L$_{\rm K_\odot} \sim 5.67 \times 10^{31}$~erg~s$^{-1}$ respectively.

\section{The sample of isolated ellipticals}

The four elliptical galaxies we discuss in this paper are part of a
larger sample of bright isolated galaxies (Focardi \& Kelm~2009), 
which we selected from the Updated Zwicky Catalog (UZC, Falco et al.~1999) 
applying an adapted version of the Focardi \& Kelm (2002) neighbour search code. 
The latter is a versatile tool
that can be applied to 3D galaxy catalogs to select galaxy samples 
characterized by different values of the 
luminosity and/or belonging to different environments (from the extreme field to galaxy clusters).
UZC is a wide angle 3D catalog of nearby galaxies that covers the 
northern sky down to a declination $\delta \sim -2.5\degr$,
and is claimed to be 96\% complete for galaxies brighter than m$_{\rm B}$ = 15.5.
The code was already successfully applied to UZC and to  the 2dFGRS
(Colless et al.~2001; Colless et al.~2003)
to extract different homogeneous samples (see 
Focardi \& Kelm 2002; Kelm et al.~2005; Focardi et al.~2006; Focardi \& Kelm~2009).

To select the sample of isolated galaxies four basic criteria were
applied to the UZC: {\bf a)} minimum B luminosity
(L$_{\rm B}$ =  1.3*10$^{10}$h$_{75}^{-2}$  L$_{\rm B\odot}$);
 {\bf b)}
velocity range (v$_{r}$  $\in$ [2500-5000] km s$^{-1}$); 
 {\bf c)}
galactic latitude ($|$b$^{\rm II}$$|$ $\ge $ 15$\degr$); and  {\bf d)} 
no companion galaxies in 3-D space
(within a given radius
R$_{iso}$=  1.3 h$_{75}^{-1}$~Mpc and radial velocity  
$|$ $\Delta$ v$_r$  $|$ = 1000 km s$^{-1}$).

The lower limit in optical luminosity  increases the chance of detecting X-ray
emission; in fact, we are interested in checking the presence of an extended, hot, 
X-ray emitting ISM, and no detectable extended halo is  expected from  
optically faint galaxies (Trinchieri \& Fabbiano 1985; Canizares et al 1987; O'Sullivan et al.~2001b).
The completeness limit
of the catalog implies a search for companions with L$_{\rm
B}$ $\ge$  5.5 $\times$ 10$^{9}$h$_{75}^{-2}$ L$_{\rm B\odot}$ (from  m$_{\rm B}$ =
15.5 at the upper limit of the radial velocity range adopted,  v$_{\rm r}$
= 5000 km s$^{-1}$).  The lower limit in radial velocity was
imposed to reduce distance uncertainties due to peculiar motions
and any contamination by the Virgo cluster, while the galactic latitude
restriction avoids an artificial increase of isolated galaxies at low
latitudes due to galactic extinction.

Isolated ellipticals
must be selected in an objective way and their isolation must be proved 
on a scale comparable to the one of galaxy groups
and clusters and to some depth (i.e. considering neighbours within a wide magnitude range).
Isolation is granted in our sample on the typical cluster/group
scale
for companions down to $\sim$
2~mag fainter and having velocity difference
well above the typical velocity dispersion
in clusters.

As a whole the sample contains 43 galaxies, among which 8 are classified
as early-type galaxies. For only four of these X-ray data exist: 
NGC~2954 and NGC~7785, for which we obtained XMM-{\em Newton} data, NGC~7052, available in the 
{\em Chandra} archive, and NGC 6172 available in the {\em ASCA} archive.
We list in Table~\ref{opt-tab} the basic optical
data for the four early-type galaxies discussed here\footnote{Note that
the recession velocity values in Table~1 are Virgo-corrected and for
this reason exceed, in two cases (NGC 6172 and NGC 7052), the original limit imposed on UZC
heliocentric radial velocities.}. The  
Virgo-corrected recession velocities and m$_{\rm B_{\rm TC}}$ used to
derive the $B$-band luminosities are from LEDA; the 
optical dimensions and  2MASS $K$-band magnitudes and extinction used to
calculate the $K$-band luminosity come  from NED
(\citealt{schlegel});
the stellar mass is estimated from the $K$-band luminosity following
the relation of Bell et al.~(2003)
and using the  B-V colors from LEDA.

Since the original selection  of the sample, new data have become
available that provide magnitudes (mostly $K$-band) and a few 
redshifts for potential companion galaxies. We checked that,
with the new data, potential companions are still at least $\sim2$~mag
fainter than our sample galaxies, confirming our original selection.
We further notice that NGC 2954 and NGC 7785 are present in the
AMIGA database (Verdes-Montenegro et al. 2005, http://amiga.iaa.es:8080/DATABASE/)
as CIG 358 and CIG 1045, and are also included in the \citet{Smith} sample.
NGC 6172 is listed as isolated in \citet{colbert} and \citet{reda}.                                            
%

\begin{table*}
\caption{\scriptsize Basic optical parameters for the isolated galaxies. 
Galaxy distances are Virgo-centric.
}
\label{opt-tab}
\begin{tabular}{cccccccc}
\hline 
\hline
\noalign{\smallskip}
Source  &  Vel. & D & Scale &Opt$_{\rm dim}$ & Log~L$_{\rm B}$ & Log~L$_{\rm K}$ & Log M$_*$ \\
 &  km s$^{-1}$ & Mpc & kpc/$\arcsec$ &$\prime$ & L$_{\rm B_\odot}$ & L$_{\rm K_\odot}$
& M$_\odot$ \\
\noalign{\smallskip}
\hline
\noalign{\smallskip}
NGC 2954 & 3774 & 51 & 0.241 & 1.1$\times$1.7 & 10.36 & 11.00 & 10.78\\
NGC 6172 & 5085 & 69 & 0.323 & 1.0$\times$1.0 & 10.46 & 11.12 & 10.92\\
NGC 7052 & 5082 & 69 & 0.323 & 1.4$\times$2.5 & 10.61 & 11.61 & 11.41\\
NGC 7785 & 3876 & 52 & 0.247 & 1.3$\times$2.5 & 10.70 & 11.42 & 11.22\\
\noalign{\smallskip}
\hline
\\
\end{tabular}
\end{table*}


\begin{figure*}
\resizebox{17cm}{!}{
\psfig{figure=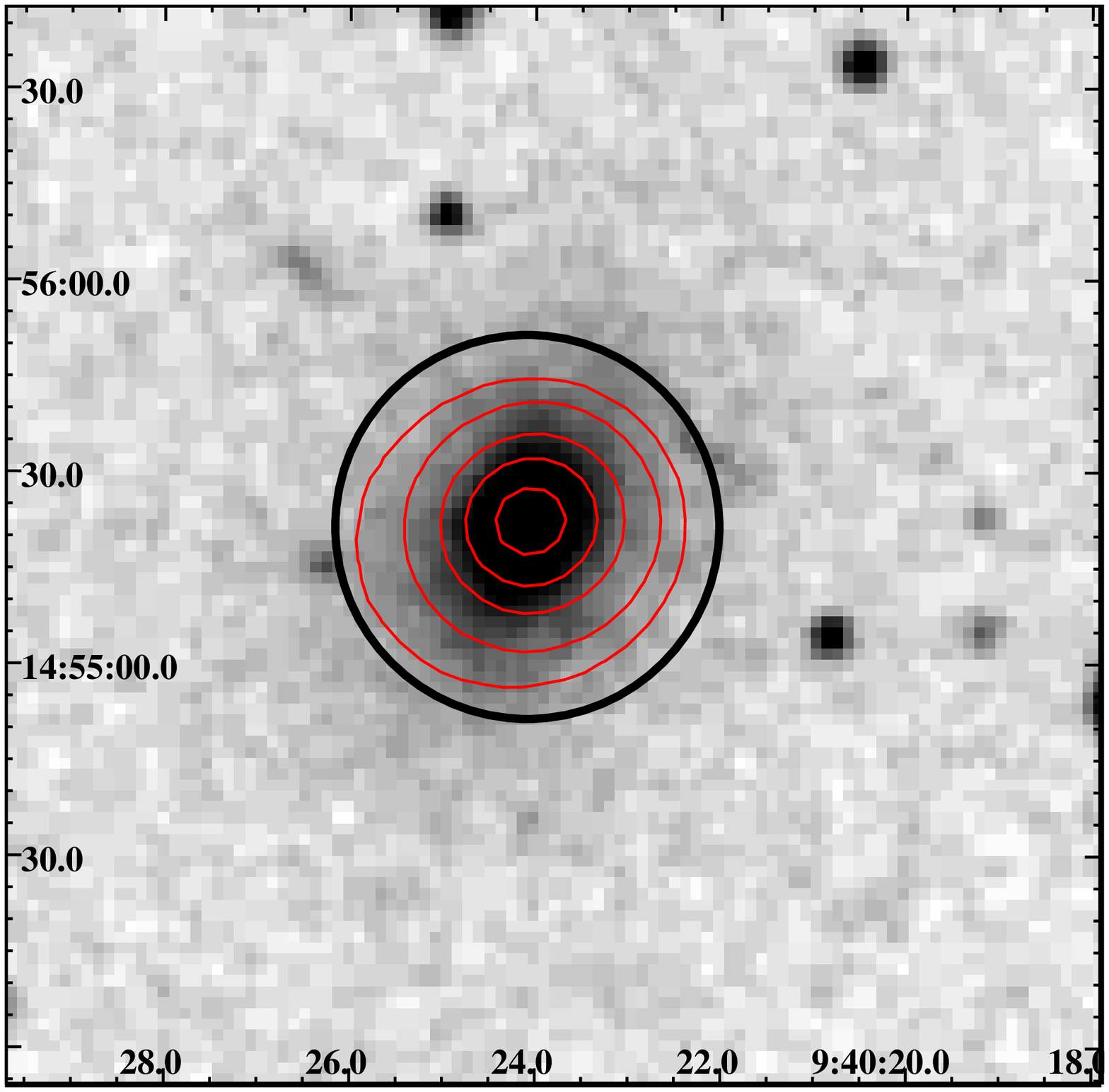,width=15.5cm}
\hspace*{1cm}
\psfig{figure=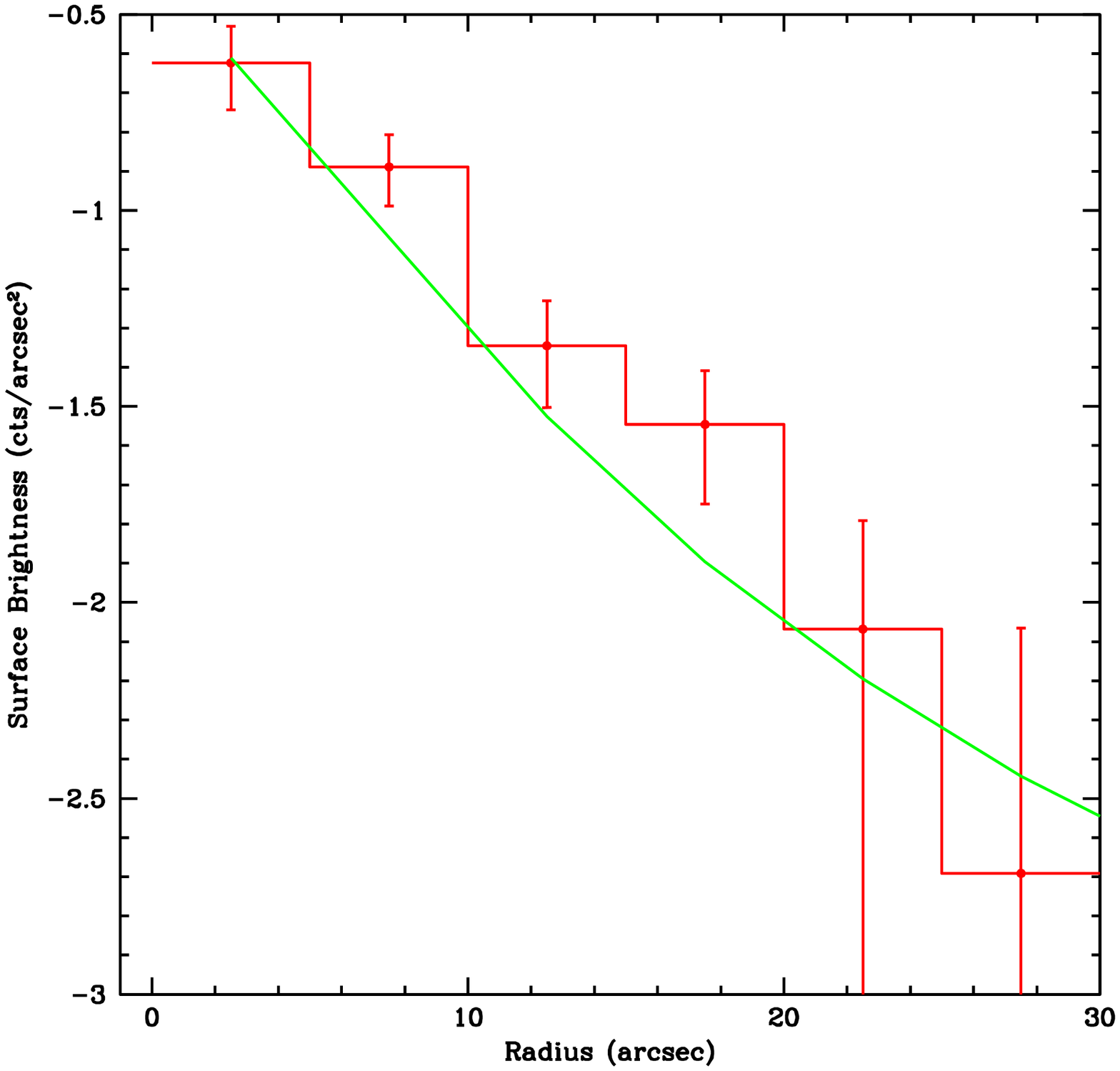,width=16.0cm}
}
\caption{\scriptsize
{\em Left panel}:  X-ray contours from the smoothed EPIC-MOS image (0.5-2.0~keV band) superposed
onto the Digitized Sky Survey image (from SAO-DSS)
for NGC 2954.
A circle of 30$''$radius, corresponding to the X-ray extraction radius used for the spectral
analysis, is shown.
{\em Right panel}: Histogram: net radial surface brightness profile centered at the X-ray peak
from the EPIC-pn data in the 0.5-2.0~keV
energy band. Continuous line: XMM-{\em Newton} Point Spread Function (see text).
\label{2954op}}
\end{figure*}

\begin{figure*}
\resizebox{17cm}{!}{
\psfig{figure=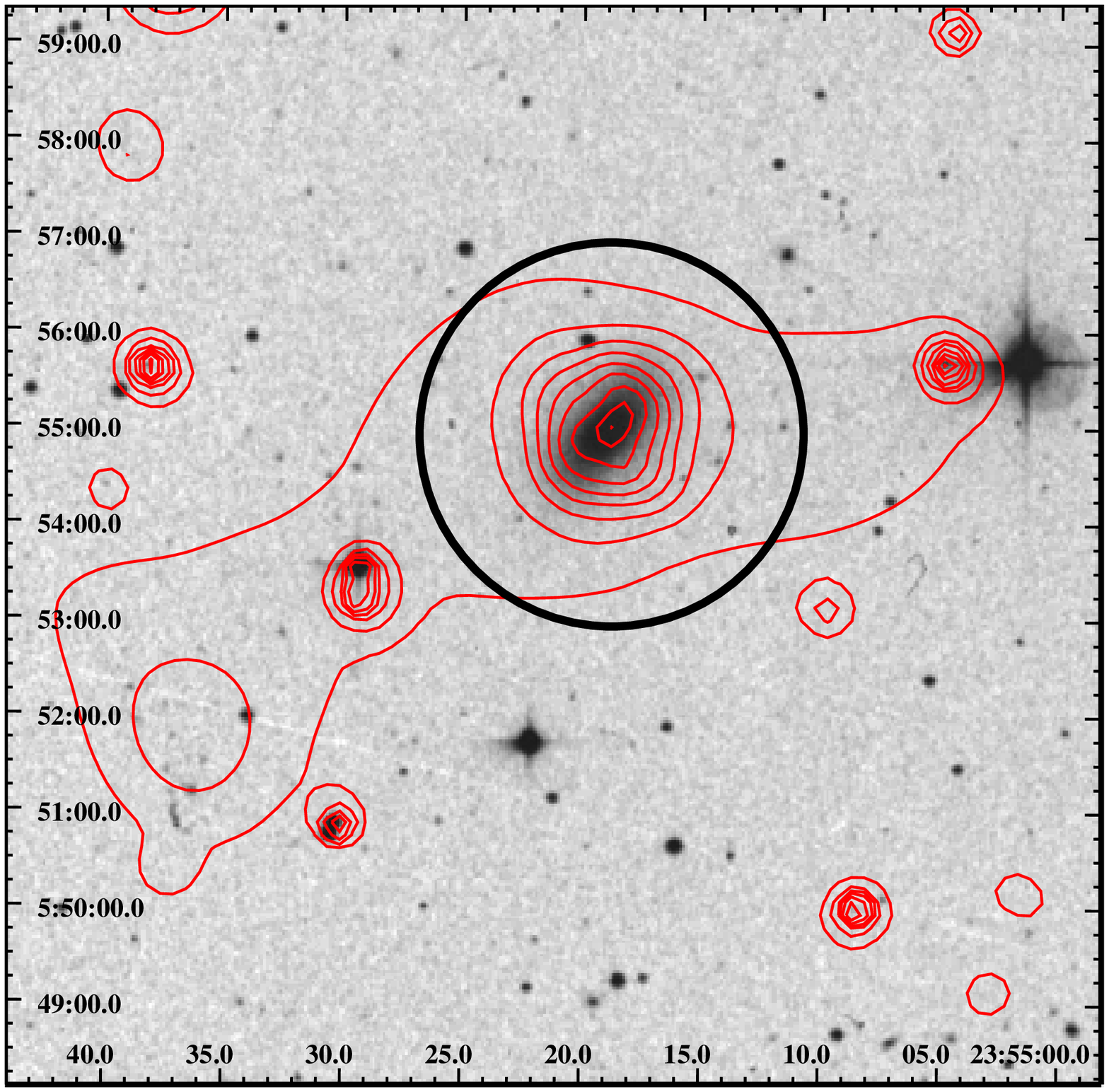,width=15.5cm}
\hspace*{0.5cm}
\psfig{figure=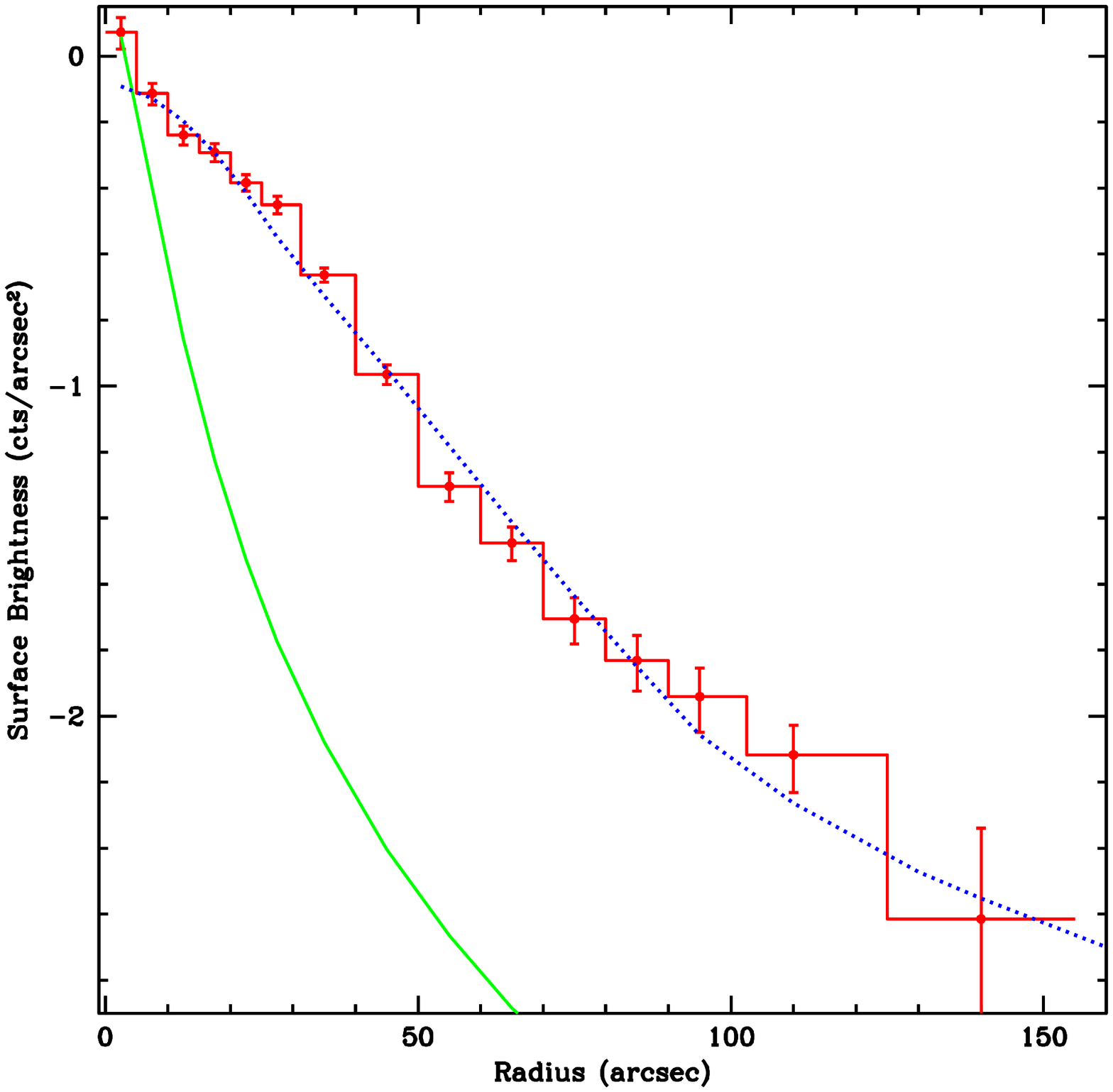,width=16.0cm}
}
\caption{\scriptsize
Same as Fig.~\ref{2954op} for NGC 7785.   
The extraction radius shown in the left panel
corresponds to $120''$. 
The dotted line represents the best fit $\beta$-model for NGC~7785 (see
text for details). 
\label{7785op}}
\end{figure*}

\begin{figure*}
\resizebox{17cm}{!}{
\psfig{figure=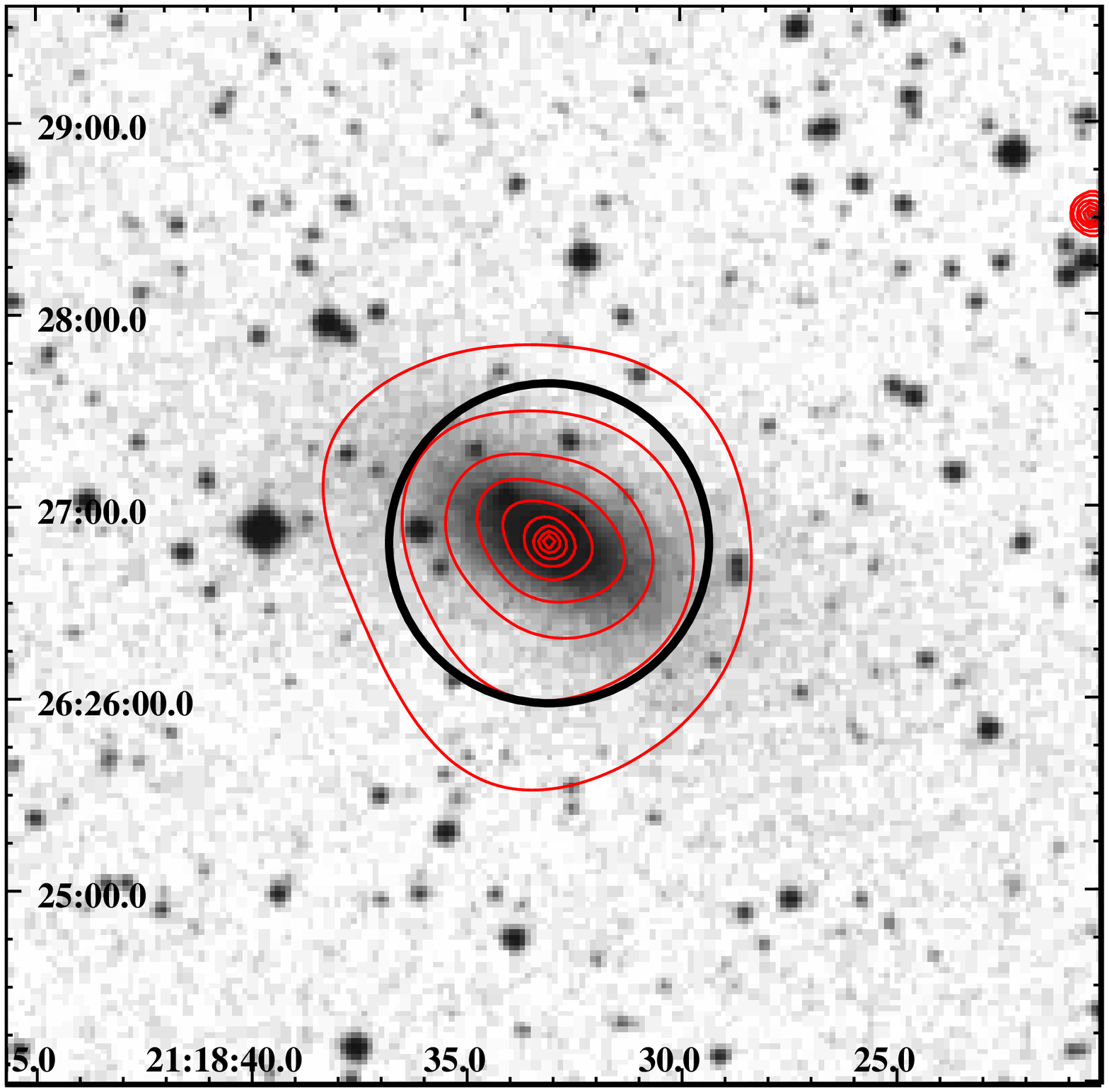,width=15.5cm}
\hspace*{0.5cm}
\psfig{figure=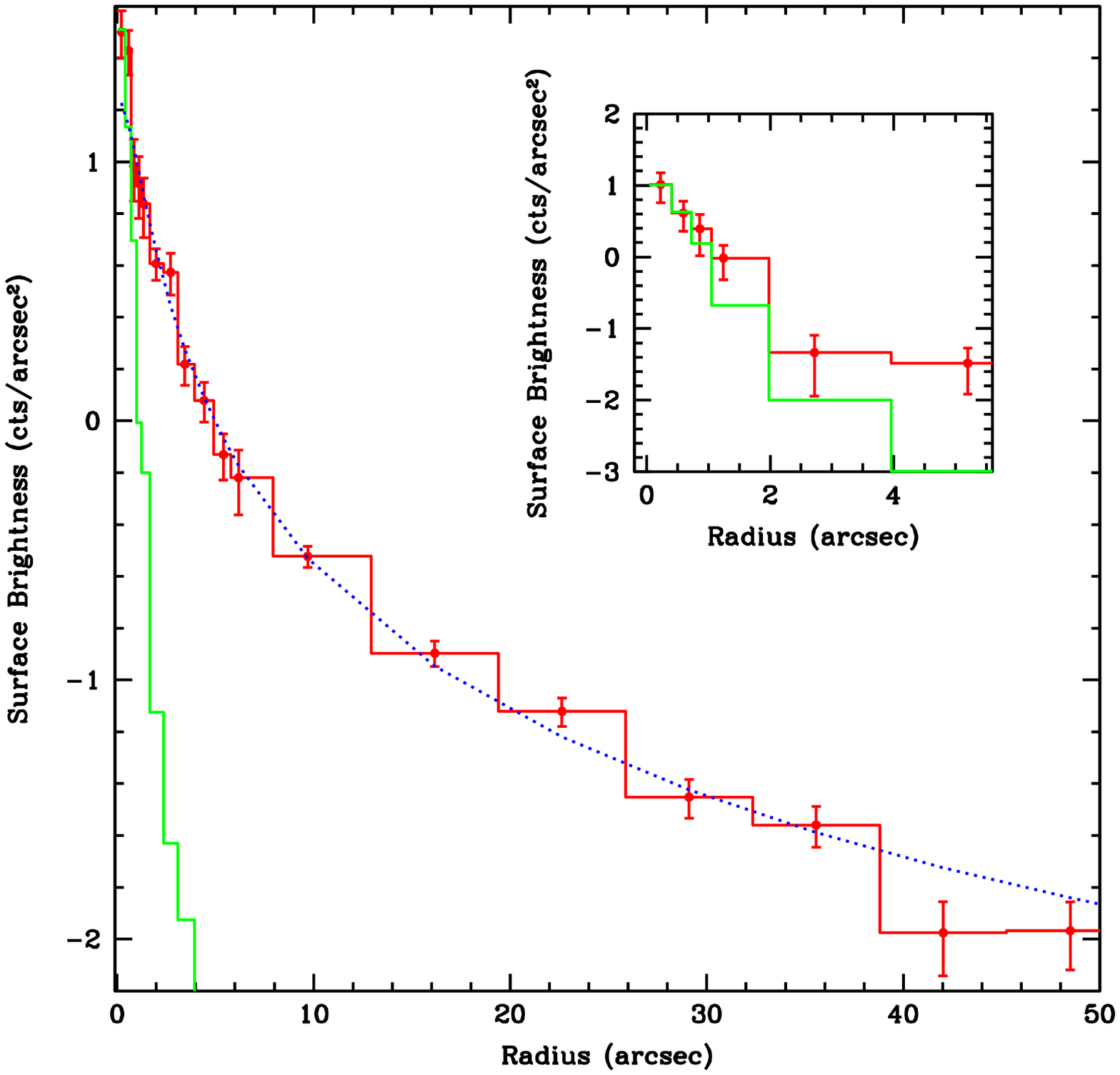,width=16.0cm}
}
\caption{\scriptsize
{\em Left panel}: X-ray contours from the smoothed {\em Chandra}-ACIS-S image (0.5-2.0~keV)
superposed onto the Digitized Sky Survey image (from SAO-DSS)
for NGC~7052.
A circle of 50$''$radius, corresponding to the X-ray extraction radius used for the spectral
analysis, is shown.
{\em Right panel}: Histogram (red): net radial surface brightness profile centered at the X-ray peak
from ACIS-S data in the 0.5-2.0~keV energy band. 
The {\em Chandra} simulated Point Spread Function is also represented with an histogram (green). 
The {\em inset} in the right panel shows a zoom of the   
net radial surface brightness profile in the 2.0-5.0~keV energy band, and the corresponding
PSF (see color figure online). 
The dotted line represents the best fit $\beta$-model for NGC~7052 (see
text for details).
\label{7052op}}
\end{figure*}

\section{X-ray observations}

\subsection{XMM-{\em Newton} data}

We obtained XMM-{\em Newton} observations for NGC 2954 and NGC 7785
between June and  November 2004, with exposure times of $\sim$30~ksec
(see Table~\ref{xmmlog} for observation details).  The EPIC detectors
operated in PrimeFullWindow mode.  We reprocessed the data with the
XMM-{\em Newton} Science Analysis Software (XMM-SAS) version 6.5.0,
using the latest calibration products.  To analyze the scientific data
we also used SAOImage DS9, FTOOLS, CIAO version 3.4, and Funtools.

We cleaned the XMM-{\em Newton} data from high flaring background periods
using the standard technique described in the User's Guide 
to the {\em XMM-Newton} Science Analysis System.
We first extracted a high-energy (10-15~keV) light-curve with time-bin=100~s
and excluded all ``flaring'' intervals 
using the standard count-rate cut (RATE$=$0.35 cts/s for EPIC-MOS
and RATE$=$1.0 cts/s for EPIC-pn). 
Taking into account the Good Time Intervals (GTIs)
and selecting single and double events (PATTERN$\leq$4) for the pn,
and single, double, triple, and quadruple events (PATTERN$\leq$12) 
for the MOS, we constructed the clean event files. 
We created 
X-ray images in different
energy bands and with different pixel-binning for the three EPIC instruments. 
We also produced mask files that correspond to bad pixels,
bad columns and CCD gaps, which were used to exclude them from the analysis. 
For the morphological studies we also merged the EPIC-MOS1 and EPIC-MOS2 event files.
Otherwise, we used the instrument files separately to obtain the spectral
photon distribution of our targets (see \S\ref{spectra}). 

\subsection{Archival data}

We searched the X-ray archives for additional observations of the 
isolated galaxies in the sample.
We found {\em Chandra} data for NGC 7052  and ASCA data for NGC 6172.
NGC 7052 was observed with the {\em Chandra} ACIS-S instrument in
September 2002.
The instrument operated in faint mode, and observed the target
for about 10~ks. 
Upon checking the data quality, we found that the event files provided
were not contaminated by high background or flaring intervals, and could
therefore be used to create images 
and extract spectral data (CIAO version 3.4), 
in analogy to what we did for the {\em XMM-Newton} data. \\
\indent
NGC 6172 was observed with   {\em ASCA} on February, 1999 for 57.2~ks.
We analyzed the data following the guidelines described in ``The {\em ASCA} 
Reduction Guide'' found at {\tt \scriptsize http://heasarc.gsfc.nasa.gov/docs/asca/abc/abc.html.}
We used a 90\arcsec\ radius source region, corresponding to 
50\% of the encircled energy fraction for the {\em ASCA} PSF,
and a wider nearby circle for estimating the background.
We derived a $\sim 1.4 \sigma$ positive signal (not sufficient to
claim a detection of the source), which corresponds to a flux F$_{{\rm
tot}(0.5-2.0)} \sim {\rm a ~~ few} ~~ 10^{-15}$ erg cm$^{-2}$ s$^{-1}$
(see Table~\ref{xmmlog}).
At the distance of the source, this corresponds to a luminosity of
L$_{{\rm tot}(0.5-2.0)} \sim 1 \times 10^{40}$ erg s$^{-1}$, a value that we will
use to estimate the maximum total luminosity of this object.

\begin{figure}
\centering
\includegraphics[width=9cm, angle=0]{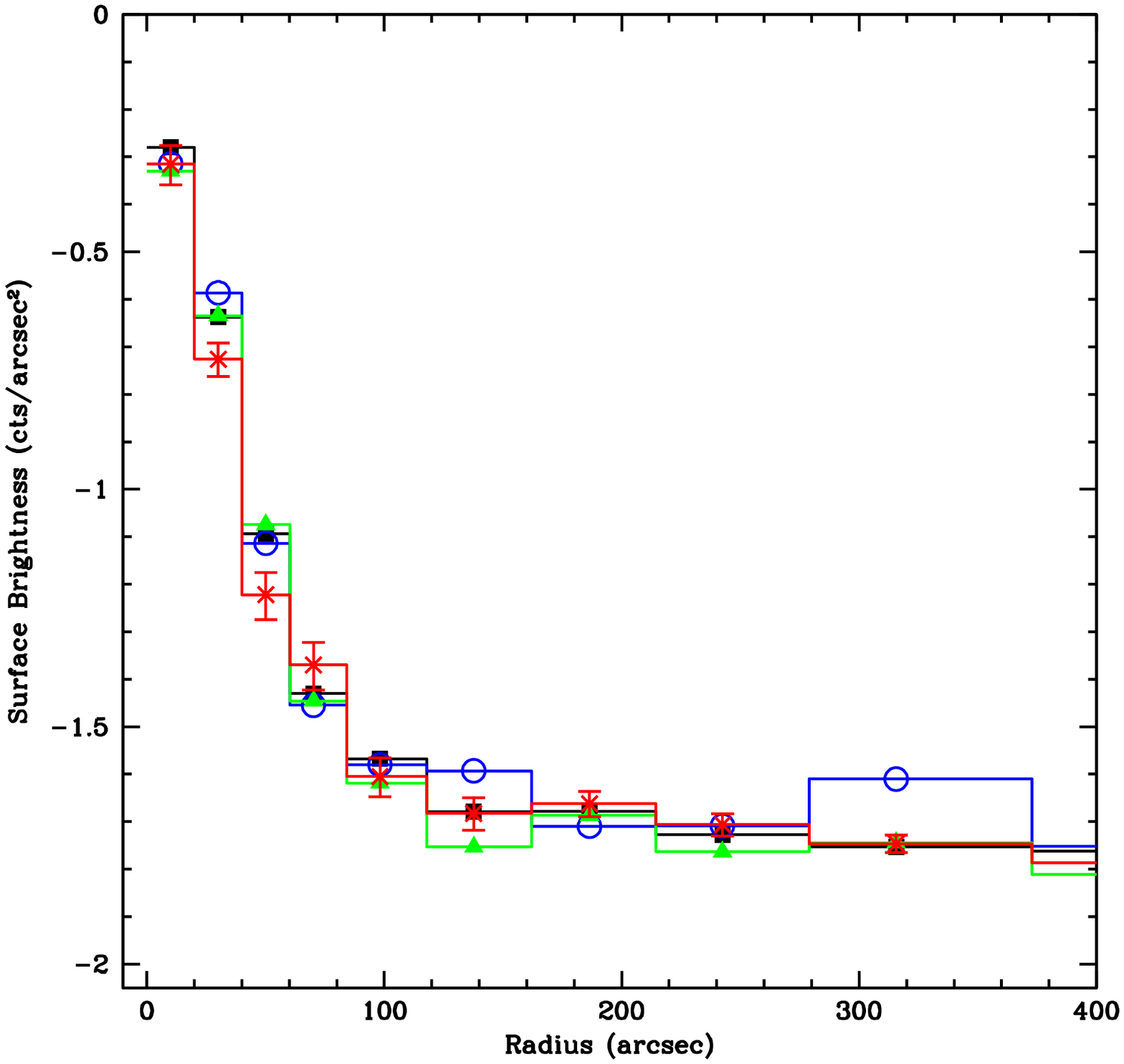}
\caption{\scriptsize 
EPIC-MOS raw surface brightness profiles
(with no background subtraction)
of NGC~7785 in the angular sectors defined in the text
(see \S~4.2):
1) N (black line and squares);  
2) SE (blue line and open circles);
3) SW (green line and triangles);
4) W (red line and crosses).
For clarity, we only report the error-bars relative to 
the Western profile (see color figure online). 
}
\label{image2}
\end{figure}

\section{X-ray spatial and spectral analysis}

\subsection{X-ray maps}

For both {\em XMM-Newton} and {\em Chandra} data-sets we produced
X-ray images with 2$''\times2''$ pixel resolution in the energy bands
0.5-2.0~keV, 2.0-4.5~keV, and 4.5-7.5~keV.  We adaptively smoothed all
of them with the {\em csmooth} tool, from CIAO, with a minimum S/N of 3.
X-ray contours from the smoothed images in the
0.5-2.0~keV band, which maximizes the S/N for all sources, are plotted
over the DSS optical images in Figs.~\ref{2954op},~\ref{7785op},
and \ref{7052op} ({\em left panels}). \\
\indent
NGC~2954 and NGC~7785 show relevant emission only in the 0.5-2.0~keV
band, whereas NGC~7052 is visible both in the 0.5-2.0~keV and in the
2.0-4.5~keV images.  
NGC 2954 and NGC 7052 are associated with compact and azimuthally symmetric
sources, centered on the optical galaxy and of similar extent. 
NGC~7785 shows a more complex morphology, with the X-ray
contours clearly elongated in East-West direction. Two compact sources
and an additional enhancement at $\ge 7'$ South-East are also visible
in the image.  Faint optical counterparts appear associated with
the two compact sources while the more diffuse  peak to the  South-East
coincides with the cluster of galaxies ZwCl 2353.2$+$0535.  It is likely
therefore that the extension is due to the presence of interlopers,
rather than to emission related to the galaxy itself.  
We will examine more quantitatively this evidence in the next section.

\subsection{Radial surface brightness profiles}
To determine the extent of the emission and the possible presence of a
nucleus, we examined
the radial profiles in  both XMM-{\em Newton} and {\em Chandra} data-sets.
We centered them on the peak of the X-ray emission
and calculated the radial surface brightness from the azimuthally averaged
distribution of concentric annuli, having excluded all CCD gaps/defects.
The maximum radius of the source is where the total profile changes slope
and becomes consistent with the expected X-ray profile of the background.
We used a circular region outside this radius, but close to the source,
to estimate the background.  The net profiles (cts/arsec$^2$)
in the 0.5-2.0 keV band are shown in Figs.~\ref{2954op},~\ref{7785op} and \ref{7052op} 
({\em right panels}), together with a comparison with the instrumental Point Spread
Function (PSF).  
We parametrized the EPIC PSF with a King-type function:
$ PSF (r) = A \, (1+(r/r_c)^2)^{\alpha}, $ where A is the normalization,
r$_{\rm c} \sim 5''- 6\farcs6$ is the core radius (MOS and pn respectively)
and $\alpha \sim -1.5$ is the slope (see the XMM-SOC-CAL-TN-0022 and
XMM-SOC-CAL-TN-0029 documents online at {\tt http://xmm.vilspa.esa.es}).
For the PSF of {\em Chandra} we used the {\em mkpsf} tool of CIAO with
the spectrum appropriate to the source (see next section).

The profile of NGC 2954 is consistent with the XMM-Newton PSF, 
and shows a positive signal only out to 30\arcsec.

Given the complex morphology of NGC~7785 described above, we analyzed
the radial profile of the emission in four different azimuthal sectors,
delimited by 90$^{\circ}$, 170$^{\circ}$, 240$^{\circ}$, and 315$^{\circ}$
counterclockwise from North.
We excluded the visible compact sources with a 25\arcsec~radius circle.
Fig.~\ref{image2} shows the raw surface brightness profiles in the angular sectors defined above.
We used EPIC-MOS images, because they are less affected by CCD gap regions. 
We confirm that no significant asymmetry in the emission is present 
from different directions and
that the emission does not extend beyond $\sim 120'' \div 160''$. 
The peak visible at $\sim 280'' \div 380''$
in the SE region is most likely associated with extended emission from the  
ZwCl 2353.2$+$0535 cluster not properly subtracted.
We then parametrized the azimuthally averaged net profile
with a King-type model, convolved with the XMM-Newton PSF described
above. We find that a model with  r$_0 \sim 35''$
and $\beta\sim0.95$ provides a good fit down to the central regions, 
consistent with no significant central excess, which could be the
signature of emission from a  nuclear source.
The derived $\beta$-model is plotted in Fig.~\ref{7785op} ({\em Right panel}).

The net profile of NGC 7052 presented in Fig.~\ref{7052op} ({\em Right panel}) 
is consistent with the broad band profile previously presented and
parametrized by \cite{donato}, as part of their sample of
FRI galaxies (NGC~7052 is the radio source B2~2116+26).  
Their best fit parametrization with a $\beta$-model 
 (r$_{\rm c} \sim 1.11''$, $\beta\sim0.48$)
superposed on the data in the {\em Right panel} of Fig.~\ref{7052op}
indicates a small central excess, consistent with their result suggesting
the presence of a point source at the 95.6$\%$ confidence.
With our different choice of energy intervals, we are able to notice 
that the central point source becomes 
evident at energies E $>$ 2~keV, 
while there is only a marginal excess over the best-fit $\beta$-model
in the 0.5-2.0 keV band.  The inset in Fig.~\ref{7052op} shows a blow up
of the inner part of the 2.0-5.0~keV profile, where the photon distribution
is entirely consistent with the Chandra PSF.   As  shown by the 0.5-2.0
keV profile in Fig.~\ref{7052op}, the total emission in NGC 7052
extends out to 50\arcsec.

\subsection{X-ray spectra}
\label{spectra}

We extracted EPIC source counts of all three instruments separately
from a circular region, 
centered on the source, with a radius of 30$''$ for NGC 2954 
and of 120$''$ for NGC 7785. Background spectra were extracted
from circular source-free regions of  $50''$ and $\sim 90''$ respectively,
close to the target. The ancillary response matrix (ARF) and the detector response matrix (RMF)
were created by means of the XMM-SAS tasks {\em {\mbox arf}gen} and {\em rmfgen}. 
The {\em Chandra} source and background spectra for NGC 7052, 
together with the appropriate spectral matrixes, were extracted using {\em specextract}
from two circular regions with a radius of 50$''$ each. 

The X-ray spectra were analyzed with the XSPEC package 
(version 11.3.1; Arnaud 1996).
EPIC-MOS1, -MOS2 and -pn spectra were fitted simultaneously.
The source counts were grouped
into energy bins such that each bin contains more than 20 counts 
to use the chi-square ($\chi^2$) statistics
and has a {\mbox signifi}cance level of at least 2$\sigma$ after background subtraction.
The quoted errors on the best-fit parameters correspond to the 90\% confidence level
for one interesting parameter 
(i.~e., $\Delta\chi^2 = 2.71$; Avni 1976).
We used a power law to model the X-ray binaries or the AGN component 
(usually parametrized by $\Gamma=1.7\div1.9$), and a {\em mekal} component, with abundances
fixed at 50\% the solar value, to fit
the diffuse hot gas. For both components, the appropriate Galactic Hydrogen column
density (N$_{\rm H}$) along the line of sight 
(Dickey \& Lockman 1990) has been taken into account. 
The spectral fitting results are summarized in Table~\ref{xmmtab}.

\begin{figure} 
\hspace*{0.05cm}
\includegraphics[width=5.7cm, angle=-90]{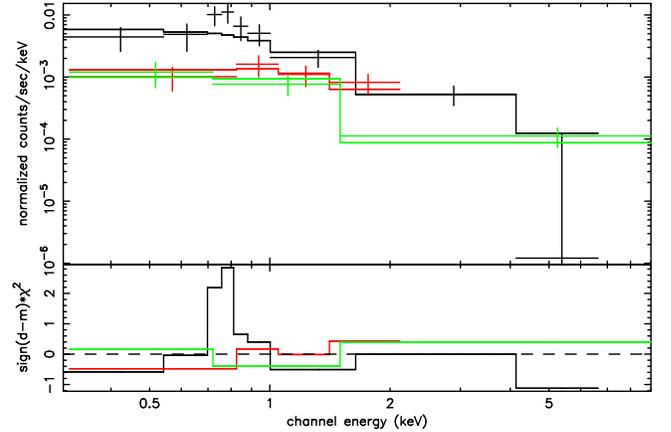}
\caption{\scriptsize NGC 2954: XMM-{\em Newton} EPIC-MOS1 (red), -MOS2 (green), and -pn (black)  
spectrum and $\chi^2$ behavior.
The best-fit model is a power law (see color figure online).}
\label{2954ld}
\end{figure}

\begin{figure*}
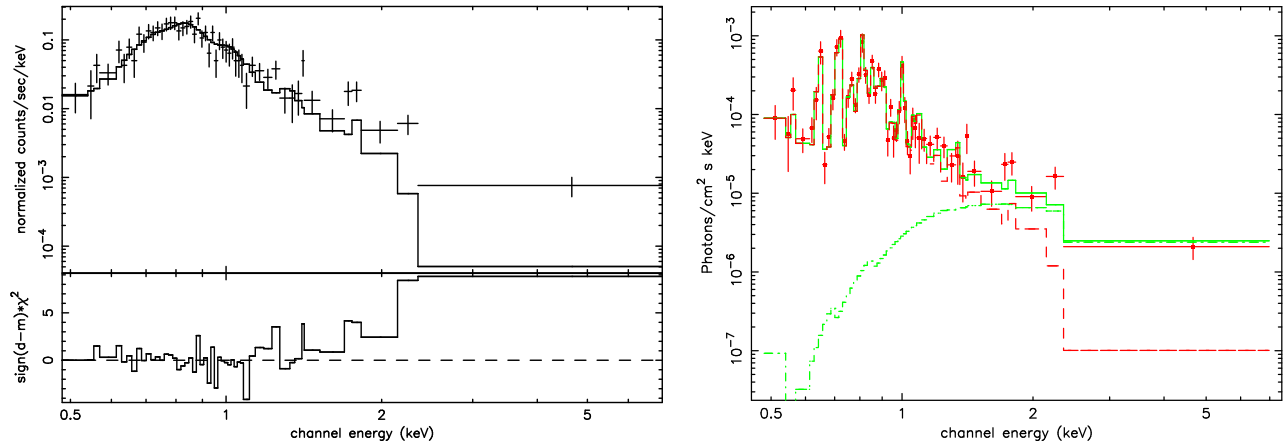
 
\resizebox{17cm}{!}{
\psfig{figure=Fig6a.ps,width=16.0cm,angle=-90}
\hspace*{1.0cm}
\psfig{figure=Fig6b.ps,width=16.0cm,angle=-90}
}
\caption{\scriptsize {\em Left panel}: {\em Chandra} ACIS-S X-ray spectrum and $\chi^2$ behavior
for NGC 7052,
fitted with a single  {\em mekal} component. 
{\em Right panel}: unfolded ACIS-S spectrum, with the best fit model composed of an absorbed power law plus
a thermal component. }
\label{7052ld}
\end{figure*}

\begin{figure*}
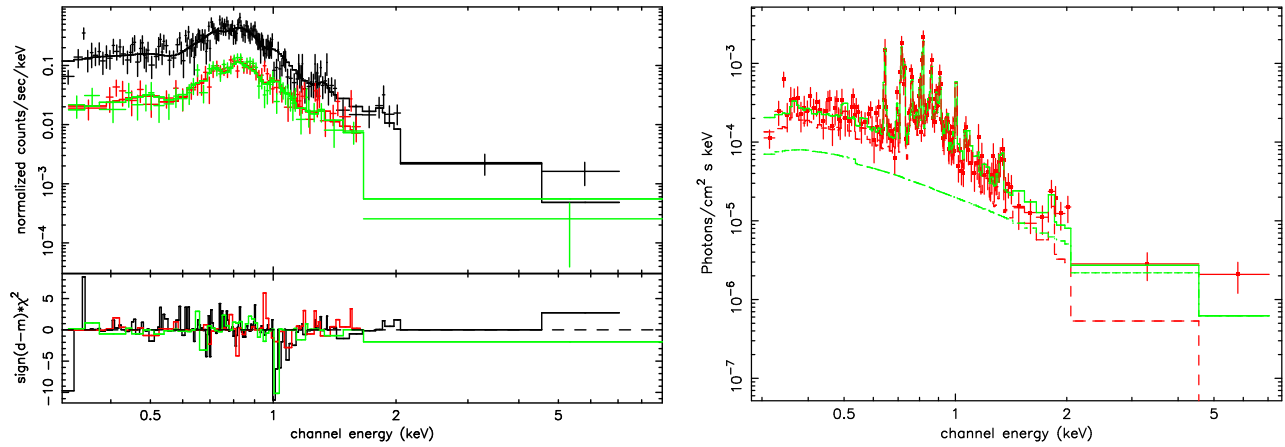
 
\resizebox{17cm}{!}{
\psfig{figure=Fig7a.ps,width=16.0cm,angle=-90}
\hspace*{1.0cm}
\psfig{figure=Fig7b.ps,width=16.0cm,angle=-90}
}
\caption{\scriptsize {\em Left panel}:
XMM-{\em Newton} EPIC-MOS1 (red), -MOS2 (green), and -pn (black)  
spectrum and $\chi^2$ behavior for NGC~7785.
{\em Right panel}: unfolded X-ray spectrum derived from the EPIC-pn data. 
The best-fit model is a power law plus a thermal component (see color figure online).}
\label{7785ld}
\end{figure*}

\begin{table*}
\begin{center}
\caption{\scriptsize Observation log.}
\label{xmmlog}
\begin{tabular}{lcccccc}
\hline \hline
\noalign{\smallskip}
Source  &  Obs.Date & ks$^{*}$ & cts/s (0.5-5.0~keV) & r$_{\rm ext}$ &Filter & Instrument \\
\noalign{\smallskip}
\hline
\noalign{\smallskip}
NGC 2954 &    2004 11 03  &  29.4/30.3/19.5 &
(1.42$\pm$0.37/1.58$\pm$0.36/6.21$\pm$0.87)$\times 10^{-3} $  & 30\arcsec &  Medium &  MOS1/MOS2/pn  \\

NGC 7785 &    2004 06 14  &  22.0/23.0/14.6 &
(4.73$\pm$0.25/4.59$\pm$0.23/19.7$\pm$0.53)$\times 10^{-2}$ & 120\arcsec&  Thin   &    MOS1/MOS2/pn  \\

NGC 7052 &    2002 09 21  &  9.63 &  (7.38$\pm$0.31)$\times 10^{-2} $   & 50\arcsec&  -- &    ACIS-S \\
NGC 6172 &    1999 02 15  & 57.2/52.1  &                --                                &
90\arcsec &  --  &    GIS/SIS  \\
\noalign{\smallskip}
\hline 
\end{tabular}
\end{center}

\begin{list}{}{}
\item[$^{*}$] {\scriptsize Live time of the central CCD after cleaning from high background flaring events
(see text).}
\end{list}
\end{table*}

\begin{table*}
\caption{\scriptsize Summary of the spectral results.
The EPIC-pn/ACIS-S net counts are derived in a circle of radius r$_{\rm ext}$, given in
the third column. 
}
\label{xmmtab}
\begin{tabular}{ccccccccccc}
\hline 
\hline
\noalign{\smallskip}
Source  &  N$_{\rm H_{gal}}$ & r$_{\rm ext}$ 
  & cts & kT$_{\rm mekal}$ & $\Gamma$ & N$_{\rm H_{int}}$  &
  $\chi^{2}/dof$ & L$_{\rm mekal
 (0.5-2.0)}$ & L$_{\rm po(2.0-10)}$ & L$_{\rm LMXBs}^{***}$ \\
 &  cm$^{-2}$ & $\arcsec$ & pn/ACIS-S & keV & & cm$^{-2}$ & & erg s$^{-1}$ & erg s$^{-1}$ & erg s$^{-1}$\\
\noalign{\smallskip}
\hline
\noalign{\smallskip}
NGC 2954       & 3.43 $\times 10^{20}$ & 30  & 132$\pm$19  &  --  &  1.9$\pm 0.3$ & -- & $10.36/12$  & -- & 3.2 $\times 10^{39}$  & 1.1 $\times 10^{40}$  \\
NGC 2954       & 3.43 $\times  10^{20}$ & 30  & 132$\pm$19  &  0.54  & 1.9$^{*}$  & -- & $8.10/11$  &  ${5.4 \times 10^{38}}^{**}$ & 2.5 $\times 10^{39}$  & 1.1 $\times 10^{40}$\\
\noalign{\smallskip}
\noalign{\smallskip}
NGC 7052            & 13.6 $\times 10^{20}$  &  50  &  719$\pm$30 &  0.48$\pm$0.06  &  1.4$^{+0.4}_{-0.8}$ &  & $47/48$ & 1.4 $\times 10^{41}$ &  5.3 $\times 10^{40}$ & 4.5 $\times 10^{40}$ \\
NGC 7052            & 13.6 $\times 10^{20}$  &  50  &  719$\pm$30 &  0.48$\pm$0.06  &  1.9$^{*}$ & $9^{+1.1}_{-0.7} \times 10^{21}$ & $46/48$ & 1 $\times 10^{41}$ &  5 $\times 10^{40}$ & 4.5 $\times 10^{40}$ \\
\noalign{\smallskip}
\noalign{\smallskip}
NGC 7785             &  5.24 $\times 10^{20}$  &  120  & 3340$\pm$87  &  0.56$\pm$0.02  & 2.0$\pm$0.3  & -- & $263/254$ & 8 $\times 10^{40}$ &  2.0 $\times 10^{40}$  & 3.0 $\times 10^{40}$ \\

\noalign{\smallskip}
\hline
\end{tabular}
\begin{list}{}{}
\item[$^{\rm *}$] {\scriptsize Fixed;
$^{\rm **}$ Note that this is not a statistically
significant component. We use it to derive an upper limit of 
L$_{\rm X}$ $ \le 1.5 \times 10^{39}$ erg s$^{-1}$  for the plasma component (see text for details);} 
\item[$^{\rm ***}$]  {\scriptsize L$_{\rm (2.0-10)}$ expected from LMXBs,
calculated using the relation in Kim \& Fabbiano (2004),
taking into account the appropriate $K$-band luminosity and energy range.} 
\end{list}
\end{table*}

\subsubsection{NGC 2954}
A simple power law  with a photon index $\Gamma$=1.9 is a good fit
and the X-ray luminosity is L$_{\rm po(2.0-10)}= 3.2 \times 10^{39}$ erg s$^{-1}$. 
However, the residuals (see Fig.~\ref{2954ld}) are not uniformly distributed,
and show an excess at $\sim$0.8~keV. If we parametrize this with 
a {\em mekal} model,
keeping the power law fixed at the best fit $\Gamma$=1.9, we derive a
gas temperature of 0.54~keV for a luminosity 
L$_{\rm mekal (0.5-2.0)} = 5.4 \times 10^{38}$ erg s$^{-1}$. 
The additional {\em mekal} component is not required by the statistics, but it allows us
to derive an upper limit on the gas contribution.
We estimated a limit with a $>95$\% confidence by raising 
the {\em mekal} component normalization
until $\Delta\chi^{2}>4$ 
(see Avni 1976).
With this normalization the upper limit value is
L$_{\rm mekal (0.5-2.0)} = 1.5 \times 10^{39}$ erg s$^{-1}$.

\subsubsection{NGC 7052}

Two components are required to fit the spectral data: 
a mekal component with a
temperature of 0.5 keV, and a power law with  photon index value $\Gamma$=1.4. 
The power-law slope appears to be flatter than expected from a population
of binaries.  
Moreover, the evidence of a central point source in the radial profile
(see also Donato et al.~2004) 
suggests that we could expect  a contribution
from a central AGN. 
In AGN, slopes as flat as  $\Gamma$=1.4 have been reported, however
they have often been interpreted as the result of a more typical power-law spectrum
(i.e.~$\Gamma$=1.9) heavily
suppressed at low energy by large absorption and poorly sampled at high energies,
due to limited
statistics and energy band 
(e.g.~Nandra \& Pounds 1995; Caccianiga et al.~2004; Piconcelli et al.~2005).
Therefore, we derive an 
alternative spectral parametrization assuming a power law with
index fixed at
$\Gamma$=1.9, which gives an intrinsic absorption 
N$_{\rm H}$ = $9^{+1.1}_{-0.7} \times 10^{21}$~cm$^{-2}$ (see
Table~\ref{xmmtab}). This gives an unabsorbed luminosity 
L$_{\rm po(2.0-10)}$ = 5 $\times 10^{40}$~erg s$^{-1}$. 
With either model for the hard energy emission, the plasma component has 
L$_{\rm mekal(0.5-2.0)} \sim 1 \times 10^{41}$~erg s$^{-1}$.
Both models are equally plausible and give the same luminosity
in the hard component. Moreover, the plasma component has also the
same luminosity of $\sim 10^{41}$ erg s$^{-1}$ in both cases. \\
\indent
In order to better define the contribution of the AGN component
separately from the one of the unresolved X-ray binaries we also extracted the
X-ray spectrum of NGC~7052 in two separate, but contiguous, regions:
a 3\arcsec\  radius circle  and a concentric annulus,
with 3\arcsec\ and 50\arcsec\ radii. 
In the inner region,  for a power law with index fixed
at $\Gamma$=1.9, the 
N$_{\rm H}$ value is consistent, within the errors, with the value
in Table~3. This component has L$_{\rm po_{(2-10)}}$ = 2.8 $\times
10^{40}$ erg s$^{-1}$, 
which we consider due to the AGN, since the binary emission
should be small due to the area considered.
Given the low statistics of the data at high energies, we cannot
easily constrain the model, therefore we assume a power law with fixed
slope  ($\Gamma$=1.9) and N$_{\rm H}$ at the galactic value
 to describe the binary emission in the outer annulus,
which results in 
an unabsorbed luminosity 
L$_{\rm po_{(2-10)}}$ = 2.2 $\times 10^{40}$ erg s$^{-1}$. \\
\indent
As already mentioned in the previous section, NGC 7052 is part of the
sample of FRI galaxies discussed by \citet{donato}.  
We are in substantial agreement with these authors.
However, given our different aims, considering
the whole source out to a radius of 50$''$, rather than to  1.5\arcsec\
as done by \citet{donato},  has provided us with significantly more
photons, allowing a better definition of
the different spectral components and a handle on the modelling of the
high energy excess. 
In fact, the request of a power-law to model the high energy part of the
spectrum is significant in our data, as shown by Fig.~\ref{7052ld}~({\em Left panel}). 
The difference in the total luminosity 
L$_{\rm (0.3-8)}$ = 4 $\times 10^{40}$ erg s$^{-1}$ calculated by \citet{donato},
with respect to our L$_{\rm (0.5-10)}$ = 15 $\times 10^{40}$ erg s$^{-1}$
is entirely due to the wider aperture we use, which includes the contribution of the ``whole'' 
galactic component (plasma and the X-ray binaries).
The estimate of the AGN luminosity is slightly higher than the upper limit
given in Donato et al.~(2004). 
This could be due to different spectral assumptions,
which they do not explicitly report.

\subsubsection{NGC 7785}
The spectrum requires both a power-law and a mekal component 
with $\Gamma$=2.0 and kT=0.56~keV, respectively (see Fig.~\ref{7785ld} and Table~\ref{xmmtab}). 
Both components are well determined: the plasma component dominates at low energies ($<2$~keV), 
with an intrinsic 
L$_{\rm mekal (0.5-2.0)}$ = 8 $\times 10^{40}$~erg s$^{-1}$ 
to be compared with L$_{\rm po(0.5-2.0)}$ = 1.4 $\times10^{40}$~erg s$^{-1}$.  
At higher energies ($>2$~keV), there
is the sole contribution of the  power-law, with L$_{\rm po(2.0-10)}$ = 2.0 $\times
10^{40}$~erg s$^{-1}$.

\section{X-ray emission from isolated galaxies}

In spite of the homogeneous selection criteria 
and environmental characteristics (see \S2)
our 4 isolated elliptical galaxies display different
X-ray properties (see \S4, Table~\ref{xmmtab}).
The total luminosities
differ considerably, from L$\rm_X > 10^{41}$ erg s$^{-1}$ for NGC
7785 and NGC 7052 to L$\rm_X \sim 3 \times 10^{39}$ erg s$^{-1}$ for NGC 2954
(0.5-10 keV).
No overall detection is obtained for NGC 6172 resulting in a total 
luminosity  L$\rm_X < 10^{40}$ erg s$^{-1}$. 

NGC~2954 has a compact
morphology consistent with the instrumental PSF, and its spectrum is well 
represented by a power-law component.  Given the lack of evidence of a nuclear source, 
we attribute this emission to the binary population.  The contribution from  binary 
sources is also detected in  NGC~7052 and 
NGC~7785, with a
luminosity of  L$\rm _{(2.0-10)} \sim
2 \times 10^{40}$~erg s$^{-1}$. 
In all three cases this contribution
is consistent with what is observed in other galaxies, based on their  L$_{\rm X}$  vs L$_{\rm K}$ ratio
\citep[see, e.g.][]{Kimfab},  as indicated by the estimated
luminosities reported in the last column of Table~\ref{xmmtab}. 

The nucleus of NGC 7052 is resolved in the radial profile and has  a luminosity of $\sim 3\times
10^{40}$~erg s$^{-1}$ (see also Donato et al.~2004).
We found no evidence of a nuclear source in the X-ray data of
NGC 7785, either in the profile or in the spectral properties.
Evidence
of an active nucleus is provided by radio emission 
(Condon et al.~2002), and its classification as a {\em core} galaxy by \citet{Lauer}, 
with a $\sim 8.5$ M$_{\sun}$ central black hole 
(van der Marel \& van den Bosch 1998).
However, the 1.4~GHz radio power of Log L$_{1.4}= 22.07$ W Hz$^{-1}$
indicates that the nucleus is
relatively faint (it is also about an order of magnitude fainter
than  NGC 7052, at Log L$_{1.4}= 23.13$ W Hz$^{-1}$; 
Condon et al.~2002),  and could very well be below the sensitivity of 
current X-ray data.  

We detect a hot ISM  in NGC~7052 and 
NGC~7785, with an extent of 16 and 30 kpc in radius, respectively, 
L$\rm_X \sim 10^{41}$ erg s$^{-1}$ (0.5-2.0 keV) and a temperature 
kT$\sim$0.5 keV.  We have also estimated a maximum contribution from a hot thin plasma component in NGC~2954, 
L$\rm_X \le 1.5 \times 10^{39}$ erg s$^{-1}$ (\S 4.3.1), which 
is significantly lower than in the other two objects. Moreover, if we compare it with the 
X-ray luminosity expected from the integrated coronal emission from the
stellar component (as estimated for example from the bulge of M31, M32, the Galactic Ridge, 
and NGC 821,
Bogd\'an \& Gilfanov 2008; Revnivtsev et al.~2007, 2006; Sazonov et al.~2006; Pellegrini et al.~2007),
we expect that the hot gas contribution is really negligible in this galaxy.
  
From the spectral parameters of the hot plasma in NGC~7052 and NGC~7785, we 
estimate a gas mass M$\rm _{gas} \sim$ 2.2  $\times 10^{9}~\rm{M}_{\odot}$
for NGC~7052 and  $\sim$ $4.6\times 10^{9}~\rm{M}_{\odot}$ for  NGC~7785,
assuming a volume corresponding to the maximum observed  radius and
spherical symmetry. 
These mass values are consistent with the 
present day stellar mass-loss rate from asymptotic giant branch
stars of $\dot{\rm M}_*$ = 0.078 (L$_{\rm B}/10^{10}$L$_{\rm B_\odot}$) M$_{\rm \odot}$ yr$^{-1}$ 
(Athey et al.~2002)
integrated over the age of the galaxies.  

Using the evidence of a hot halo, we can also estimate the total mass of these two galaxies.
To calculate the total mass we have used the expression in Fabricant et al.~(1984),
derived under the  assumption  that the gas is isothermal in hydrostatic equilibrium 
in the potential, worked out explicitly in terms of $\beta$ and $R_{core}$ 
by  Mulchaey \& Zabludoff (1999). 
Applying their Eq.~(3) to NGC 7052 and NGC 7785 with the appropriate parameters,
we derive M$\rm _{tot} \sim 5 \times 10^{11}$ M$_\odot$ out to a radius of  16 kpc 
for NGC 7052 (M/L = 12.3 M$_{\odot}$/L$_{\rm B_\odot}$)
and M$_{\rm tot} \sim 1.9\times 10^{12}$ M$_\odot$ out to a radius of 30 kpc
for NGC 7785 (M/L = 37.9 M$_{\odot}$/L$_{\rm B_\odot}$).

Comparing them with the estimates of the visible mass 
(see Table~1), these values indicate a large dark matter halo only in NGC 7785 
(about an order of magnitude larger than the stellar mass). 
While we have no  support from the data that the assumption
of hydrostatic equilibrium is correct, and therefore the total masses derived 
from the X-ray data could be incorrect, nevertheless we note that NGC 7785 has a 
large M/L ratio derived independently of the stellar velocity dispersion 
(Heckman 1983).
We cannot derive the total mass for the other two objects from the X-ray data.
However, Heckman (1983) gives a mass for NGC 2954 of M$ \sim 5\times 10^{11}$ M$_\odot$  within the 
effective radius (normalized to our H$_0$ value), which also suggests a large amount of 
dark matter in this galaxy.  

\begin{figure}
\includegraphics[width=9cm, angle=0]{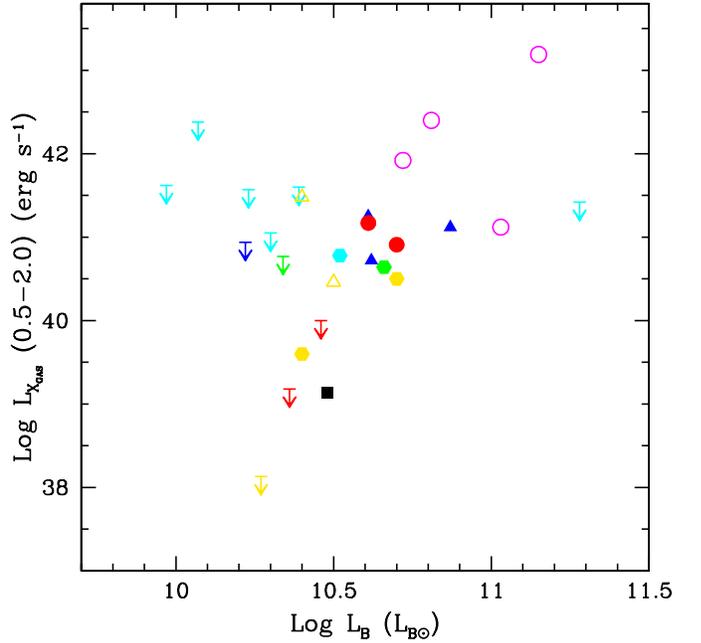}
\caption{\scriptsize 
Scatter plot of the 0.5-2.0~keV hot gas X-ray luminosity vs $B$-band luminosity
of 23 early-type isolated galaxies and 4 isolated early-type pairs
(see color figure online).
L$_{\rm X}$ refer to the gas component contribution in 0.5-2.0~keV except 
the 4 sources plotted as
{\em hexagons}, for which only the total L$_{\rm X}$ is available. 
 Arrows indicate upper limits.
{\em Red dots and arrows}: this paper;
{\em blue triangles}: O'Sullivan et al.~(2007);
{\em blue arrow}: O'Sullivan et al.~(2001b);
{\em green hexagon and arrow}:  Helsdon et al.~(2001) with no companion within 1.0~Mpc;
{\em cyan hexagon and arrows}: Helsdon et al.~(2001) with no companion within 0.4~Mpc;
{\em magenta open circles}: ``fossil'' group ellipticals from O'Sullivan et al.~(2001b), 
and  Mulchaey \& Zabludoff (1999), and ``fossil'' cluster with data from Fukazawa et al.~(2006);
{\em black square}: Fukazawa et al.~(2006);
{\em yellow hexagons and open triangles}: Gr\"utzbauch et al.~(2007);
{\em yellow arrow}:  Pellegrini et al.~(2007).
}
\label{image1}
\end{figure}

To better estimate the diversity of the X-ray properties of the isolated
galaxies, in particular in their ISM content, 
we compare them in Fig.~\ref{image1}, where  
we plot the L$_{\rm X}$ of the gas vs L$_{\rm B}$. 
In the narrow range of optical luminosity, to within a factor of 3, the
X-ray luminosity of the ISM spans more than 2 orders of magnitudes, in particular
if we consider that the upper limit to the ``plasma'' luminosity estimated for NGC 2954
could in fact be significantly lower due to a significant contribution expected 
from the stellar component.   As discussed above, 
the total mass does not appear to be a better tracer of a hot ISM, even based on just the 
estimates for NGC 2954 and NGC 7785.

To better populate the region, and to exclude that the scatter is due to
the very small sample used, we have searched the literature for isolated early-type galaxies
for which a measure of the gas component is available.  We 
plot 23 additional objects  in Fig.~\ref{image1}. 
The galaxies are selected with  isolation criteria similar to ours,
from the sample of O'Sullivan et al.~(2001b, 2007: ESO107-4,
NGC~57, IC1531, and NGC7796), and Helsdon et al.~(2001:
NGC~6776 and NGC~2271, plus seven galaxies with a somewhat less stringent
isolation radius of 0.4~Mpc, mostly upper limits), plus 
NGC 2865 (Fukazawa et al.~2006), and NGC~821 (Pellegrini et al.~2007).  We also include four
``fossil'' groups (NGC 6482 and ESO306-17: O'Sullivan et al.~2001b [note that the total mass 
of NGC 6482 could be significantly lower than typical fossil groups, see Buote et al.~2007];
NGC~1132: Mulchaey \& Zabludoff 1999; and NGC~1600: Fukazawa
et al.~2006), which would also pass our isolation criteria,
and 4 isolated early-type pairs (Gr\"utzbauch et al.~2007). 

The distribution of galaxies in the plot
of Fig.~\ref{image1} indicates that the two bands cover very different luminosity ranges 
and there is a notable lack of optically bright and X-ray faint objects. 
The first is nothing but a different representation of what has been known now
for a few decades, namely that the hot  gas properties appear to be
relatively independent of the optical ones.  We merely use a well
defined sample of isolated galaxies, therefore supporting the evidence
that this (lack of) relation is intrinsic to this class of sources and
not induced by environmental effects.  
We need instead to better understand whether the second effect is due to
a selection bias.  The sample we use in this work is made of galaxies
of relatively low luminosity. 
This could be related to an evidence of  a general 
lack of luminous isolated early-type galaxies in the local universe
(e.g.~Kelm \& Focardi 2004; Sulentic et al.~2006),
but it could also be a consequence of the small
volume probed.  In fact a work by
Stocke et al.~(2004)  indicates that galaxies forming in poorly populated  
environments are  just as likely to become  very luminous elliptical
galaxies as those in dense groups and in rich clusters.
We note that, as shown by Fig.~\ref{image1} itself, 
a small relaxing of the search parameters in the volume of the
universe probed (in recession velocity or declination, rather than in the isolation 
criteria) has already provided us with a few brighter objects. 

The galaxies plotted in Fig.~\ref{image1} do not represent a complete 
sample, therefore the lack of objects in the lower right corner can be simply due to 
the fact that we have not included any.  There is no observational restriction to populate it, since current instruments have the sensitivity to
detect X-ray luminosities below the $\sim 10^{40}$ erg s$^{-1}$ limit
in a very large volume. However, neither the 
large sample used by  O'Sullivan et al.~ 
(2001b; 401 early-type galaxies with ROSAT data
selected regardless of environments, and  
possibly biased towards  X-ray bright objects),
nor the low X-ray luminosity objects,
selected for this specific property (O'Sullivan \& Ponman 2004; David et al.~2006),
appear to populate this region. 

While not conclusive, this evidence suggests that bright galaxies in general
 do retain their hot ISM.  We can further speculate 
that an optical luminosity  
Log L$_{\rm B} \sim 10.5$ L$_{\rm {B}_{_\odot}}$ 
could be used as a discriminant between two groups, with 
galaxies below such value unable to retain a hot halo, which instead, 
when present, is at the level of $10^{40}$ erg s$^{-1}$  or above in optically brighter objects.  
A possible counterexample is given by the low-luminosity (L$_{\rm B} \leq 3
\times 10^{10}$ L$_{\rm {B}_{_\odot}}$) early-type
galaxies from David et al (2006), in which hot gas is detected in most of the 18 galaxies investigated
at L${\rm_X}$ (0.5-2.0~keV)  = $2 \times 10^{38} - 2 \times 10^{40}$ erg s$^{-1}$.
However, the sample includes objects in different environments, which could modify their intrinsic properties (in fact the brightest object,
NGC 4552, is in Virgo and it is currently being stripped of its gas content, Machacek et al.~2006), and lower luminosity objects appear to be in a wind phase, 
which would therefore limit any gas accumulation (David et al.~2006).  Moreover,  at the lower X-ray luminosity, the exact contribution from hot gas might be difficult to assess with data of limited quality, 
due to both the  contribution from stellar sources and the possible presence of outflowing gas. This is  
exemplified by NGC~3379,  a very well studied nearby elliptical galaxy with deep Chandra observations (Brassington et al.~2008): as discussed by Revnivtsev et al.~(2008), most of the soft diffuse emission is well explained by the stellar population.  This is also the only galaxy in which  gas in outflow has been detected, but its contribution is very low, at L$_{\rm gas} \sim 4 \times 10^{37}$ erg s$^{-1}$, 
M$_{\rm gas} \sim (3 \pm 1) \times 10^5$ M$_{\odot}$  (Trinchieri et al.~2008b), 
well below previous estimates for a gas component.  
It is to be hoped that more detailed studies of these objects will become available, 
to help us understand better whether all isolated 
galaxies at low optical luminosities 
contain as little gas as speculated here.

\section{Comparison with theoretical expectations}

Theoretical models developed by Ciotti et al.~(1991) and Pellegrini \& Ciotti (1998) discuss different yields in the X-ray band as a function of several input parameters,  linked to the stellar and total mass  and their distribution, 
and the SN Ia rate.  Two specific examples are given for systems with  
L$_{\rm B} = 5 \times 10^{10}$ L$_{\rm {B}_{_\odot}}$ and  M$_{*} = 1.9 \times 10^{11}$ M$_\odot$,
and  L$_{\rm B} = 10^{11}$ L$_{\rm {B}_{_\odot}}$ and  M$_{*} = 4.2
\times 10^{11}$ M$_\odot$ (H$_{0}$=50~km~s$^{-1}$~Mpc$^{-1}$). In the
former example the galaxies are able to develop partial to global winds
that make the X-ray luminosity in the gas component drop to very low
values. At larger masses, even in the partial wind regime, the galaxies
would have a wider central inflow region  due to the deeper potential
well, with resulting higher gas temperatures and  L$_{\rm X}$ in the
range $3\times 10^{40}$ erg s$^{-1}$ to $2\times 10^{41}$ erg s$^{-1}$.
These values roughly correspond to our crude evaluation of two regimes
discussed above.  In this very simple scenario, NGC 7052 and NGC 7785
would be able to retain their hot ISM to levels consistent with the
integrated stellar mass loss (see previous section), while the lack of
gas in NGC 2954 and NGC 6172 could be explained as being due to a global
wind that has cleaned them of the gas shed by the stars.   Global winds
are hard to detect since they are expected to produce very low X-ray
luminosities, beyond the sensitivity and quality of our observations,
as already discussed in the previous section.

Similarly, David et al.~(2006) suggest that supernova heating is enough 
to drive a galactic wind in systems with  L$_{\rm K} < 10^{11}$ L$_{\rm {K}_{_\odot}}$, even without resorting to the energy input from a central AGN.  
Consistently, they find that the low luminosity objects they 
analyze are also relatively under-luminous in their gaseous component, 
with gas masses below the expected accumulation from the stellar evolution. 
 In spite of the general trend of having  low X-ray luminosities, their sample spans a wide range 
(about a factor $\sim 100$)  in the $\rm L_X/L_K$ ratios. 
David et al.~(2006) interpret this as possibly due to the influence of AGN heating 
and/or environment (which is usually a major concern in most discussions).
Since we have selected galaxies to be isolated, the wide range in luminosity ratios we find should not result from the role played by the environment.  

The role of AGN feedback is often invoked to explain many galaxy properties,
like for instance
the tight scaling relations between Super Massive Black Hole masses
and the properties of host galaxies' bulges (e.g.~Ferrarese \& Merrit 2000;
Marconi \& Hunt 2003), but the details of the mechanisms are not
well know yet. Cavities, bubbles and weak shocks generated by the jet in the ISM
demonstrate the ability of the AGN to deposit large amounts of
energy in the environment (most evident in clusters, but now visible in galaxies as well, 
e.g.~ Finoguenov \& Jones 2001;  Biller et al.~2004; Croton et al.~2006; Birzan et al.~2008).
While all galaxies are most likely hosting a central black hole, its 
activity can actually produce two opposite effects: a
``negative'' feedback, when gas is ejected by the ignition of a nucleus
into the inter-galactic medium, thus reducing or even
stopping star formation (Silk \& Rees 1998; Granato et al.~2001); or a
``positive'' feedback, when the compression of the inter-stellar
medium from the jets induces star-formation 
(Klamer et al.~2004).
Although, based on the data we have, we cannot claim that the  AGN 
should be held responsible for the observed scatter, 
we  notice that the two objects with evidence of AGN activity in our sample 
(i.e. radio emission) are 
also those with a sizable plasma component. A more systematic comparison of the 
radio and gas properties in isolated galaxies should be made, to better
assess the role of AGN feedback, also in light of the observed
correlation between the asymmetries in the gas morphology  and  AGN
activity in a sample of well studied objects (Diehl \& Statler 2008),
which are however in a variety of environments.

\begin{figure}
\centering
\includegraphics[width=9cm, angle=0]{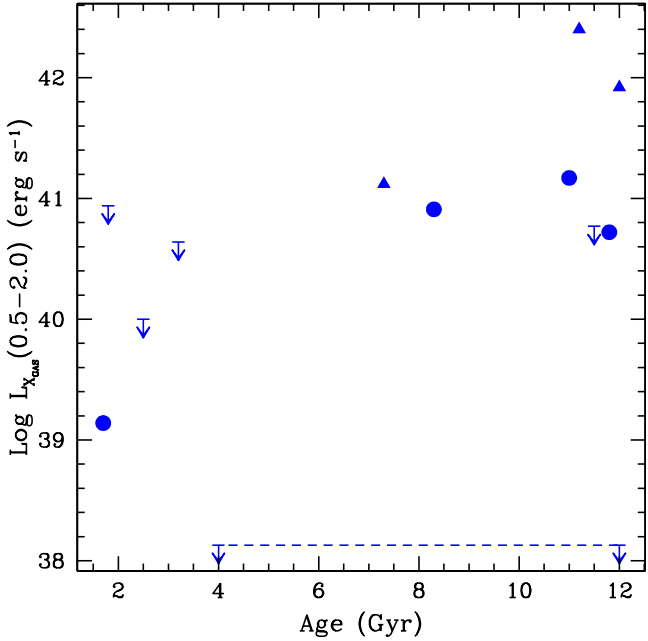}
\caption{\scriptsize Scatter plot of the 0.5-2.0~keV hot gas
X-ray luminosity vs age estimates for the isolated galaxies considered in Fig.~\ref{image1} (see \S6).
Fossil groups are plotted as triangles.
}
\label{image3}
\end{figure}

\section{The L$_{\rm X}$ vs Age relation}

An attractive explanation of the large intrinsic scatter in the
L$_{\rm X}$/L$_{\rm B}$ relation of early-type galaxies
could be found if we associate it with their merging histories.  A
first systematic attempt to correlate  the X-ray properties with ``age''
was discussed by 
Sansom et al.~(2000; see also Mackie \& Fabbiano 1997)
who noticed a strong deficiency of hot ISM in ``young'' galaxies, based on a 
sample of 69 E and S0 galaxies plus 2
merger remnants in a wide range of fine-structure index $\Sigma$, 
suggested as an indicator of previous mergers (Schweizer \& Seitzer 1992). 
Five of the sample galaxies are members of the Virgo
Cluster, while all the others are either field objects or in loose
groups. A refined discussion of X-ray emission versus age is given by Sansom et al.~(2006) 
where spectroscopic or dynamical ages are used, instead of indirect measures such as
fine structure. 
More isolated examples confirm that merger remnants are under-luminous in X
rays compared with typical mature elliptical galaxies into which these
are expected to evolve (Fabbiano \& Schweizer 1995;  Nolan et al.~2004).  Similarly, in a small sample
of pairs located in low density environments, shown in Fig.~\ref{image1},
diffuse hot gas appears to be associated with relaxed objects, while the two systems
with
unambiguous signatures of ongoing interaction  have lower total X-ray emission, indicating 
a significantly smaller contribution from a hot ISM  
(Trinchieri \& Rampazzo 2001; Gr\"utzbauch et al.~2007).

Due to the paucity of information mainly on ages but also on gas content, most work
concentrating on the ``age-gas content'' relation include all early-type galaxies for which 
information exists, regardless of the environmental properties that 
could artificially increase the
scatter in the plot, and is not negligible at almost all ages 
(see Fig.~\ref{image1} and \ref{image3} in Sansom et al.~2006). 
Here we make an attempt at separating galaxies according to their environment, and
compare the 
ISM properties as a function of age in our sample of isolated galaxies.
Unfortunately very few estimates of ages are available.  Moreover, it is well known that
ages are difficult to interpret properly, since it is often hard to break
 the age/metallicity degeneracy (see e.g.~Trager et al.~1998, 2000; Terlevich \& Forbes
2002), and that  ages obtained with different indicators might not be comparable. 
Therefore, any 
interpretation of trends with age should be considered carefully.

We found an age of 11~Gyr
and  8.3~Gyr for NGC~7052 and NGC~7785, respectively (Terlevich \& Forbes
2002), and 2.5$\pm$0.5~Gyr for  NGC~6172 (Reda et al.~2007). 
We could find no estimate for the age of  NGC~2954. If we assume an age 
of $\sim$3~Gyr for this object, which would be consistent with  the predicted trend of B--K color 
with age for a population of solar metallicity 
(see, e.g.~Fig.~\ref{2954ld} of Gallagher et al.~2008)\footnote{We are fully aware that this is a rather arbitrary assumption,
given the spread of observed B--K color in galaxies at almost any age $>$ 3 Gyr in the same plot.},
 we notice that 
the two brighter and gas rich galaxies appear to be ``old" ($>8$~Gyr),
while the low luminosity systems appear to be younger  ($<$ 3 Gyr). 

Again we resort to a search in the literature to enrich the sample, 
and we add in Fig.~\ref{image3} the isolated galaxies considered above 
for which an age estimate is available 
(NGC 7796: Thomas et al.~2005; NGC 1132, NGC 2271, and NGC2865: Reda et al.~2007;
NGC 6482: Humphrey et al.~2006;
NGC 6776, NGC 1600, and ESO107-4:  Terlevich \& Forbes 2002; NGC 821: 
Denicol\'o et al.~2005 and Proctor et al.~2005).
The plot  shows a significantly smaller scatter than previous attempts 
(O'Sullivan et al.~2001a; Samson et al.~2006; Trinchieri et al.~2008a) and seems to indeed suggest that age could be a key factor:  young systems, at ages $\le
 3$ Gyr, are systematically faint (in fact only NGC 2865 is  detected). 
``Old'' systems, with ages $>7$ Gyr, appear instead to have a measurable gas component. 
If we exclude NGC 1132 and NGC 6482 
(which are in fact fossil groups, with a significantly larger total mass than our objects),
old objects appear to have a gas luminosity of L$\rm_X \sim 10^{41}$ erg s$^{-1}$. 
Young shell galaxies, which are
typically  in a poor environment though not as isolated as the galaxies here, also appear to
follow this trend and show a negligible gas content (Trinchieri et al.~2008a).

However, we have a strong case in NGC 821 that could dramatically add to the scatter:
its low gas content would suggest a young age (e.g.~4.0~Gyr,  Denicol\'o et
al.~2005), however most estimates indicate an older system (7.7~Gyr,
Trager et al.~2000; 11.5~Gyr, Caldwell et al.~2003; 
12.5 Gyr, Vazdekis et al.~2004).
In particular, Pellegrini et al.~(2007) adopt an age of 12~Gyr
(Proctor et al.~2005), and through 
hydrodynamical simulations specific for NGC~821
suggest that the hot gas is driven out of the galaxy in a
wind sustained by type Ia supernovae, although even with their deep observations they are not able to actually detect it.
If the age of NGC 821 is indeed greater than 7 Gyr, than this
object alone provides a scatter of about 2 orders of magnitude in the L$\rm_X$ of older objects.
 
A few attempts at explaining the suggested trend of  X-ray luminosity with age have been put forward,  and involve both infall of gas or accumulation from the stellar mass loss, after the merging event that has 
depleted the system from its gas content.  
Given the extremely poor environment of these systems, infall, either as primordial HI 
(Hibbard \& van Gorkom 1996) or as hot gas previously  
being driven out as a result of the merging event, would require a massive dark matter halo to prevent the gas from abandoning the  system. 
O'Sullivan et al.~(2001a) argue against infall, which would predict a time-scale for the halo regeneration 
inconsistent with the observations along the merging sequence, and favour a declining rate in type Ia 
supernovae that would consequentially reduce the energy deposition in the ISM with time, and produce  a
transition from outflow (characteristic of younger objects) to inflow and  accumulation
of the gas shed by stars.  Among the old detected objects, NGC 1132 (Mulchaey \& Zabludoff 1999), 
NGC 6482 (Buote et al.~2007),
NGC 1600 (Smith et al.~2008),
and  NGC 7785 are all relatively massive systems, while  NGC 7052 and  NGC 7796 (O'Sullivan et al.~2007)
appear to be less massive.  Except for NGC 1132, the measured gas mass is also consistent with the expected accumulation at the current mass loss rate 
(see \S5 and Sivakoff et al.~2004; O'Sullivan et al.~2007),
which, in this context, would indicate that all  systems are massive
enough to be able to retain most of the regenerated halo.  We have no
detailed information on current supernova rate in these systems, to be
able to support the \cite{su07} suggestion of a decline with age.
However, we notice that NGC 821 appears to be a significantly less
massive system, consistent with it being in an outflow phase (Pellegrini
et al.~2007).
Interestingly, the observed wind in NGC 3379 is also found in an old
system with no sizable dark matter halo (see discussion in Trinchieri et
al.~2008b).  This would point again to mass as one of the key parameters,
although, as indicated by NGC 2954 (massive but of unknown age!), mass
alone does not seem to be enough to guarantee a sizable hot ISM.

\section{Conclusions}

We have presented the spatial and spectral characteristics of the X-ray
emission from four isolated early-type galaxies, selected with strict
isolation criteria in the local universe.  We find a relatively large
scatter in their X-ray properties: two of them have a sizable hot ISM,
with L$\rm_X \sim 10^{41}$ erg s$^{-1}$, while the other two appear to be at
least a factor 10 fainter and relatively free of hot gas.  A range of
X-ray properties was already reported in a few isolated galaxies  
(e.g.~O'Sullivan et al.~2007).
In fact, from our systematic attempt to compare  
the isolated objects available in the literature, we  confirm the large scatter in the relation between the X-ray characteristics 
of their ISM  and the stellar luminosity/mass, suggesting that the scatter is not induced by the environment, but it is an intrinsic property of this class of sources. 
Based on three of our objects, it would also appear that not even the total mass is a discriminant  to guarantee the presence of a hot ISM. 

Attempts at linking the origin of the scatter with other known galactic properties have not led us to conclusive results. The attractive explanation 
of relating the ISM content to age appears to be only partially successful: ``younger'' objects are typically fainter and do not appear to have a sizable
hot ISM. Older systems are brighter in their ISM content, however, if we consider the well studied system NGC 821, their X-ray luminosities could span 
over 2 orders of magnitudes.  Understanding the role of the central AGN will require better data than available at the present time.

We are planning to continue our study of the intrinsic properties of early-type galaxies with new X-ray observations and with high quality optical spectra,
to obtain a better handle on the age and the metallicity of the
stellar populations for a larger set of objects, which in turn 
will give us insights on possible different evolutions in systems with diverse X-ray properties.  

\begin{acknowledgements}
We acknowledge financial contribution from the contract ASI-INAF I/023/05/0.
We thank the anonymous referee for useful comments and
many of our colleagues for interesting discussions 
during the preparation of this work,
in particular: Stefano Andreon, Jay Gallagher, Angela Iovino, and  
Emanuela Pompei.
E.M. is grateful to Lucia Ballo, Lea Giordano, Arturo Mignano, Mara 
Salvato, Paola Severgnini, Marzia Tajer, 
and Luca Zappacosta for their scientific support throughout.
As one of thousands of Italian researchers with a short-term 
position, E.M. acknowledges the support of Nature (455, 835-836 and 840-841) 
and thanks the editors for increasing international awareness of the current
critical situation of Italian research. \\
This research made use of SAOImage DS9, and {\em funtools} developed by the 
Smithsonian Astrophysical Observatory (SAO), 
of the SAO/NASA Astrophysics Data System (ADS), 
operated by the SAO under a NASA grant,
of the NASA/IPAC Extragalactic Database (NED), operated by the Jet Propulsion Laboratory, 
California Institute of Technology, under contract with 
the National Aeronautics and Space Administration,
and of the Lyon-Meudon Extragalactic DAtabase (LEDA), created in 1983 at Lyon Observatory,
and, since that time, continuously updated.
Both CIAO and XMM-SAS softwares have been used to reduce the data. This publication made also use of 
data from the Two Micron All Sky Survey that is a joint project of the University of Massachusetts
and the Infrared Processing and Analysis Center/California Institute of Technology, funded by the 
National Aeronautics and Space Administration and the National Science Foundation. 
We also used the Digitized Sky Survey that was produced at the Space
Telescope Science Institute under US Government grant NAG W-2166,
and is based on photographic data of the National Geographic Society -- Palomar
Observatory Sky Survey (NGS-POSS) obtained using the Oschin Telescope on
Palomar Mountain.  The NGS-POSS was funded by a grant from the National
Geographic Society to the California Institute of Technology.  The
plates were processed into the present compressed digital form with
their {\mbox permission.}  
\end{acknowledgements}


\begin{thebibliography}{}

\bibitem[Andreon(2006)]{Andreon} Andreon, S.~2006, A\&A, 448, 447 

\bibitem[Arnaud(1996)]{arnaud} Arnaud, K.~A.~1996, ASPC, 101, 17

\bibitem[Athey et al.(2002)]{Athey} Athey, A., Bregman, J., Bregman, J., Temi, P., Sauvage, M.~2002, ApJ,
571, 272

\bibitem[Avni(1976)]{avni} Avni, Y.~1976, 210, 642

\bibitem[Bell et al.(2003)]{Bell} Bell, E.F., Mc Intosh D.H., \& Weinberg, M.D.~2003, ApJS, 149, 289

\bibitem[Bernardi et al.(2003)]{Bernardi} Bernardi, M., Sheth, R.K., Annis J.~et al.~2003, AJ, 125, 1882

\bibitem[Beuing et al.(1999)]{Beuing} Beuing, J., Dobereiner, S., Bohringer, H., \& Bender, R.~1999, MNRAS,
302, 209

\bibitem[Biller et al.(2004)]{Biller} Biller, B.A., Jones, C., Forman, W.R., Kraft, R., \& Ensslin, T.~2004,
ApJ, 613, 238

\bibitem[Birzan et al.(2008)]{Birzan} Birzan, L., McNamara, B.R., Nulsen, P.E.J., Carilli, C.L., \& Wise,
M.W.~2008, ApJ, 686, 859 

\bibitem[Bogd\'an \& Gilfanov(2008)]{M31} Bogd\'an, \'A., \& Gilfanov, M.~2008, MNRAS, 388, 56

\bibitem[Brassington et al.(2007)]{brassington07} Brassington, N.J., Ponman, T.J, \&
Read, A.M.~2007, MNRAS, 377, 1439

\bibitem[Brassington et al.(2008)]{brassington08} Brassington, N.J., et al.~2008, ApJS, 179, 142

\bibitem[Brown \& Bregman(1998)]{Brown} Brown, B.A., \& Bregman, J.N.~1998, ApJ, 495, L75

\bibitem[Buote et al.(2007)]{Buote} Buote, D.~A., 
Gastaldello, F., Humphrey, P.~J., Zappacosta, L., Bullock, J.~S., 
Brighenti, F., \& Mathews, W.~G.\ 2007, \apj, 664, 123 

\bibitem[Caccianiga et al.(2004)]{caccia} Caccianiga, A., Severgnini, P., Braito,
V., et al.~2004, 416, 901

\bibitem[Caldwell at al.(2003)]{Caldwell} Caldwell, N., Rose, J.A, \& Concannon K.D.~2003, AJ, 125, 2891

\bibitem[Canizares et al.(1983)]{Canizares83} Canizares C.R., Stewart, G.C. \& Fabian, A.C.~1983, ApJ, 272, 449

\bibitem[Canizares et al.(1987)]{Canizares87} Canizares, C.R., Fabbiano, G., \& Trinchieri, G.~1987, ApJ, 312, 503

\bibitem[Chiosi\& Carraro(2002)]{Chiosi} Chiosi, C., \& Carraro, G.~2002, MNRAS, 335, 335 

\bibitem[Ciotti et al.(1991)]{Ciotti} Ciotti, L., D'Ercole, A., Pellegrini, S., \& Renzini, A.~1991, ApJ,
376, 380

\bibitem[Colbert et al.(2001)]{colbert} Colbert, J.W., Mulchaey, J.S., \& Zabludoff
A.I.~2001, AJ, 121, 808 

\bibitem[Colless et al.(2001)]{Colless01} Colless, M.M., and the 2dFGRS team 2001, MNRAS, 328, 1039

\bibitem[Colless et al.(2003)]{Colless03} Colless, M.M., and the 2dFGRS team 2003yCat., 7226

\bibitem[Condon et al.(2002)]{Condond02} Condon, J.J., et al.~2002, yCat, 8065

\bibitem[Croton et al.(2006)]{Croton} Croton, D.J., Springel, V., White, S.D., et al.~2006, MNRAS, 365, 11

\bibitem[David et al.(2006)]{david} David, L.P., Jones, C., Forman, W., Vargas,
I.M., \& Nulsen, P.~2006, ApJ, 653, 207

\bibitem[De Lucia et al.(2008)]{Delucia} De Lucia, G., \& Helmi A.~2008, MNRAS, 391, 14

\bibitem[Denicol\'o et al.(2005)]{Denicolo} Denicol\'o, G., Terlevich, R., Terlevich, R., Forbes D.A., \&
Terlevich, A.~2005, MNRAS, 358, 813

\bibitem[De Propris et al.(2007)]{DePropris} De Propris, R., Conselice, C.J., Liske, J., et al.~2007, ApJ, 666, 212

\bibitem[Dickey \& Lockman(1990)]{dickey} Dickey, J.M., \& Lockman, F.J.~1990,
ARA\&A, 28, 215

\bibitem[Diehl \& Statler(2008)]{Diehl} Diehl, S., Statler, T.~2008, ApJ, 680, 897

\bibitem[Donato et al.(2004)]{donato} Donato, D., Sambruna, M.N., \& Gliozzi,
M.~2004, ApJ, 617, 915 

\bibitem[Dressler(1980)]{Dressler} Dressler, A.~1980, ApJ, 236, 351

\bibitem[Eggen et al.(1962)]{Eggen} Eggen, O.J., Lynden-Bell, D., \& Sandage, A.R.~1962, ApJ, 136, 748

\bibitem[Ellis et al.(1997)]{Ellis} Ellis, R.S., Smail, I., Dressler, A., et al.~1997, ApJ, 483, 582

\bibitem[Eskridge et al.(1995)]{Eskridge} Eskridge, P.B., Fabbiano, G., \& Kim, D.W.~1995, ApJS, 97, 141

\bibitem[Fabbiano \& Trinchieri(1985)]{Fabbiano} Fabbiano, G., \& Trinchieri, G.~1985, ApJ, 296, 430 

\bibitem[Fabbiano \& Schweizer(1995)]{Fabbiano95}  Fabbiano, G., \& Schweizer, F. 1995, ApJ, 447, 572

\bibitem[Fabricant et al.(1984)]{Fabricant} Fabricant, D., Rybicki, G.B, \& Gorestein, P.~1984, ApJ, 286, 186

\bibitem[Falco et al.(1999)]{falco} Falco, E.E., Kurtz, M.J., Geller, M.J., et
al.~1999, PASP, 111, 438

\bibitem[Ferrarese \& Merritt(200)]{Ferrarese} Ferrarese. L., \& Merritt D.~2000, ApJ, 539, L9

\bibitem[Finoguenov \& Jones(2001)]{Finoguenov} Finoguenov, A., \& Jones, C.~2001, ApJ, 547, L107

\bibitem[Focardi \& Kelm(2002)]{focardi} Focardi, P., \& Kelm, B.~2002, A\&A, 391, 35

\bibitem[Focardi et al.(2006)]{Focardi} Focardi, P., Zitelli, V., Marinoni, S., \& Kelm, B.~2006, A\&A, 456, 467

\bibitem[Focardi \& Kelm.(2009)]{Focardi} Focardi, P., \& Kelm, B.~2009, in preparation

\bibitem[Forman et al.(1985)]{Forman} Forman, W., Jones, C., \& Tucker, W.~1985, ApJ, 293, 102

\bibitem[Fukazawa et al.(2006)]{fukazawa} Fukazawa, Y., Botoya-Nonesa, J.G., Pu,
J., Ohto, A., \& Kawano, N.~2006, ApJ, 636, 698

\bibitem[Gallagher et al.(2008)]{Gallagher} Gallagher, J.S., et al.~2008, ApJ, 685, 752

\bibitem[Granato et al.(2001)]{Granato} Granato, G.L., et al.~2001, MNRAS, 324, 757

\bibitem[Gr\"utzbauch et al.(2007)]{Grutzbauch} Gr\"utzbauch, R., Trinchieri, G., Rampazzo, R., et al.~2007, AJ, 133, 220

\bibitem[Gunn |& gott(1972)]{Gunn} Gunn, J.E., \& Gott, J.R.III 1972, ApJ, 176, 1

\bibitem[Heckman(1983)]{Heckman} Heckman, T.M. 1983, ApJ, 273, 505

\bibitem[Helsdon et al.(2001)]{helsdon01} Helsdon, S.F., Ponman, T.J., O'Sullivan,
E., \& Forbes, D.A.~2001, MNRAS, 325, 693

\bibitem[Hibbard \& van Gorkom(1996)]{hibbard} Hibbard J.E.,\& van Gorkom J.H.~1996, AJ, 111, 655

\bibitem[Humphrey et al.(2006)]{humphrey06} Humphrey, P.J., et al.~2006, ApJ, 646, 899

\bibitem[Kelm \& Focardi(2004)]{Kelm04} Kelm, B., \& Focardi P.~2004, A\&A, 418, 937

\bibitem[Kelm et al.(2005)]{Kelm05} Kelm, B., Focardi, P., \& Sorrentino G.~2005, A\&A, 442, 117

\bibitem[Kim et al.(1992)]{KFT} Kim, D.W., Fabbiano, G., \& Trinchieri, G.~1992, ApJ, 393, 134

\bibitem[Kim \& Fabbiano(2004)]{Kimfab} Kim, D.W., \& Fabbiano G.~2004, ApJ, 611, 846

\bibitem[Klamer et al.(2004)]{Klamer} Klamer, I. J., Ekers, R. D., Sadler, E. M., Hunstead, R.W. 2004, ApJ, 612, L97

\bibitem[Knochfar \& Silk(2006)]{Knochfar} Knochfar, S., \& Silk J.~2006, ApJ, 648, L21 

\bibitem[Larson(1975)]{Larson} Larson 1975, MNRAS, 173, 671

\bibitem[Lauer et al.(2007)]{Lauer} Lauer, T.R., et al.~2007, ApJ, 662, 808

\bibitem[Machacek et al.(2006)]{machacek} Machacek, M., Jones, C., Forman, W.R., \& Nulsen, P.~2006, ApJ, 644, 155 

\bibitem[Mackie \& Fabbiano(1997)]{Mackie} Mackie, G., \& Fabbiano, G.~1997, ASPC, 116, 401

\bibitem[Marconi \& Hunt(2003)]{Marconi} Marconi, A., Hunt, L.K.~2003, ApJ, 589, L21

\bibitem[Marcum et al.(2004)]{Marcum} Marcum, P.M, Aars, C.E., \& Fanelli M.N.~2004, AJ, 127, 3213

\bibitem[Mathews et al.(2006)]{Mathews} Mathews, W.G., et al.~2006, ApJ, 652, L17

\bibitem[Melnick \& Sargent(1997)]{Melnick} Melnick, J., \& Sargent, W.L.W.~1977, ApJ, 215, 401

\bibitem[Mulchaey et al.(1999)]{Mulchaey99} Mulchaey, J.S., \& Zabludoff, A.I.~1999, ApJ, 514, 133

\bibitem[Nandra \& Pounds(1994)]{Nandra} Nandra, K., \& Pounds K.A.~1994, MNRAS, 268, 405

\bibitem[Nolan et al.(2004)]{Nolan} Nolan, L.A., Ponman, T.J., Read, A.M., Schweizer, F.~2004, MNRAS, 353, 221

\bibitem[Nulsen(1982)]{Nulsen} Nulsen 1982, MNRAS, 198, 1007

\bibitem[O'Sullivan et al.(2001a)]{su01a} O'Sullivan E., Forbes D.A., \& Ponman T.J.~2001a, MNRAS, 324, 420

\bibitem[O'Sullivan et al.(2001b)]{su01b} O'Sullivan E., Forbes D.A., \& Ponman T.J.~2001b, MNRAS, 328, 461

\bibitem[O'Sullivan et al.(2004)]{su04} O'Sullivan, E., \& Ponman, T.J.~2004, MNRAS, 349, 535

\bibitem[O'Sullivan et al.(2007)]{su07} O'Sullivan, E., Sanderson, A.J.R., Ponman, T.J.~2007, MNRAS, 380, 1409

\bibitem[Pellegrini \& Ciotti(1998)]{PellegriniCiotti} Pellegrini, S., Ciotti, L.~1998, A\&A, 333, 433

\bibitem[Pellegrini et al.(2007)]{Pellegrini07} Pellegrini, S., et al.~2007a, ApJ, 667, 731 

\bibitem[Piconcelli et al.(2005)]{pico} Piconcelli, E., Guainazzi, M., Cappi, M.,
Jimenez-Bailon, E., \& Schartel, N.~2005, A\&A, 432, 15

\bibitem[Proctor et al.(2005)]{Proctor} Proctor, R.N., Forbes, D.A., Forestell, A., Gebhardt, K.~2005, MNRAS,
362, 857

\bibitem[Reda et al.(2004)]{reda} Reda, F.M., Forbes, D.A., Beasley, M.A.,
O'Sullivan, E.J. \& Goudfrooij, P.~2004, MNRAS, 354, 851

\bibitem[Reda et al.(2007)]{Reda07} Reda, F.M., Proctor, R.N., Forbes, D.A., Hau, G.K.T., Larsen,
S.S.~2007, MNRAS, 377, 1772

\bibitem[Revnivtsev et al.(2006)]{Revnivtsev06} Revnivtsev, M., Sazonov, S., Gilfanov, M., Churazov, E., \& Sunyaev, R.~2006, A\&A, 452, 169

\bibitem[Revnivtsev et al.(2007)]{Revnivtsev07} Revnivtsev, M., Churazov, E., Sazonov S., Forman, W., \& Jones, C.~2007, A\&A, 473, 783

\bibitem[Revnivtsev et al.(2008)]{Revnivtsev08} Revnivtsev, M.,  Churazov, E.,
Sazonov, S., Forman, W., \& Jones, C.\ 2008,  A\&A, 490, 37

\bibitem[Sansom et al.(2000)]{Sansom00} Sansom, A.E., Hibbard, J.E., \& Schweizer, F.~2000, AJ, 120, 1946

\bibitem[Sansom et al.(2006)]{Sansom06} Sansom, A.E., O'Sullivan, E., Forbes, D.A., Proctor, R.N., \& Davis,
D.S.~2006, MNRAS, 370, 1541 

\bibitem[Sazonov et al.(2006)]{Sazonov} Sazonov, S., Revnivtsev, M., Gilfanov, M., Churazov, E., \& Sunyaev, R. 2006, A\&A, 450, 117

\bibitem[Schlegel et al.(1998)]{schlegel} Schlegel, D.J., Finkbeiner, D.P., \&
Davis, M.~1998, ApJ, 500, 525

\bibitem[Schweizer \& Seitzer(1992)]{Schweizer} Schweizer, F., \& Seitzer, P.~1992, IAUS, 149, 488

\bibitem[Silk \& Rees(1998)]{Silk} Silk, J., Rees, M.J.~1998, A\&A, 331L, 1

\bibitem[Sivakoff et al.(2004)]{siva} Sivakoff, G. R., Sarazin, C. L., \& Carlin, J. L. 2004, ApJ, 617, 262

\bibitem[Smith et al.(2004)]{Smith} Smith, R.M., Mart\'inez, V.J.,\& Graham M.J.~2004, ApJ, 617, 1017
 
\bibitem[Smith et al.(2008)]{Smith08} Smith, R.~M., 
Mart{\'{\i}}nez, V.~J., Fern{\'a}ndez-Soto, A., Ballesteros, F.~J., 
\& Ortiz-Gil, A.\ 2008, \apj, 679, 420 

\bibitem[Stocke et al.(2004)]{Stocke} Stocke, J.T., Keeney, B.A., Lewis, A.D., Epps, H.W., Schild, R.E.~2004,
AJ, 127, 1336

\bibitem[Sulentic et al.(2006)]{sulentic} Sulentic J.W., et al.~2006, A\&A, 449, 937

\bibitem[Terlevich \& Forbes(2002)]{Terlevich} Terlevich, A.I., \& Forbes, D.A.~2002, MNRAS, 330, 547 

\bibitem[Thomas et al.(2005)]{thomas} Thomas, D., Maraston, C., Bender, R., Mendes de Oliveira, C. 2005, ApJ, 621, 673

\bibitem[Tonry \& Davis(1981)]{Tonry} Tonry, J.L, \& Davis, M.~1981, ApJ, 246, 666

\bibitem[Toomre \& Toomre(1972)]{toomre} Toomre, A., \& Toomre, J.~1972, ApJ, 187, 623

\bibitem[Trager et al.(1998)]{Trager98} Trager, S.C., Worthey, G., Faber, S.M., Burstein, D., \& Gonz\'alez,
J.J.~1998, ApJS, 116, 1

\bibitem[Trager et al.(2000)]{Trager} Trager, S.C., Faber, S.M., Worthey, G., Gonz\'alez, J.J.~2000, AJ, 119, 1645

\bibitem[Treu et al.(2002)]{Treu} Treu, T., Stiavelli, M., Casertano, S., Moller, P., Bertin, G.~2002,
ApJ, 564, L13 

\bibitem[Trinchieri \& Fabbiano(1985)]{TrinchieriFabbiano} Trinchieri G., \& Fabbiano G.~1985, ApJ, 296, 447

\bibitem[Trinchieri \& Rampazzo(2001)]{TrinchieriRampazzo} Trinchieri, G., \& Rampazzo, R.~2001, A\&A, 374, 454

\bibitem[Trinchieri et al.(2008a)]{trinchieri08a} Trinchieri G., et al.~2008a, A\&A, 489, 85

\bibitem[Trinchieri et al.(2008b)]{trinchieri08b} Trinchieri G., et al.~2008b, ApJ, 688, 1000

\bibitem[Tully(1987)]{Tully} Tully, R.B.~1987, ApJ, 321, 280

\bibitem[van der Marel \& van den Bosch(1998)]{vanderMarel}van der Marel, R. P., \& van den Bosch, F. C. 1998, AJ, 116, 2220

\bibitem[van Dokkum(2005)]{vanDokkum} van Dokkum, P.G.~2005, ApJ, 130, 2647

\bibitem[Vazdekis et al.(2004)]{Vazdekis} Vazdekis, A., Trujillo, I., \& Yamada, Y.~2004, ApJ, 601, L33 

\bibitem[Verdes-Montenegro et al.(2005)]{verdes} Verdes-Montenegro, L., et al.~2005,
A\&A, 436, 443

\end{thebibliography}
\end{document}